 \documentclass[twocolumn]{aastex63}

\title{2009MS9 Paper}

\usepackage{xcolor}
\usepackage{url}
\usepackage{longtable}
\usepackage{graphicx}
\usepackage{threeparttable}
\usepackage{amsmath}

\received{January 04, 2022}
\revised{October 14, 2022}
\accepted{October 25, 2022}

\submitjournal{Planetary Science Journal}

\shorttitle{2009 MS9}
\shortauthors{Bufanda et al.}

\begin{document}

\title{TNO or Comet? The Search for Activity and Characterization of Distant Object 418993 (2009 MS9)}

\correspondingauthor{Erica Bufanda}
\email{ebufanda@hawaii.edu}
\author[0000-0002-0406-8518]{Erica Bufanda}
\affiliation{Institute for Astronomy, 2680 Woodlawn Drive, Honolulu, HI 96822 USA}

\author[0000-0002-2058-5670]{Karen J. Meech}
\affiliation{Institute for Astronomy, 2680 Woodlawn Drive, Honolulu, HI 96822 USA}

\author[0000-0002-4734-8878]{Jan T. Kleyna}
\affiliation{Institute for Astronomy, 2680 Woodlawn Drive, Honolulu, HI 96822 USA}

\author[0000-0001-6952-9349]{Olivier R. Hainaut}
\affiliation{European Southern Observatory, Karl-Schwarzschild-Strasse 2, D-85748 Garching bei M\"unchen, Germany}

\author[0000-0001-9542-0953]{James M. Bauer}
\affiliation{Department of Astronomy, University of Maryland, College Park, MD 20742-2421}

\author[0000-0003-0425-5265]{Haynes Stephens}
\affiliation{University of California, Berkeley, USA}

\author[0000-0002-5396-946X]{Peter Veres}
\affiliation{Harvard-Smithsonian Center for Astrophysics, 60 Garden St., MS-15, Cambridge, MA 02138, USA}

\author[0000-0001-7895-8209]{Marco Micheli}
\affiliation{ESA SSA-NEO Coordination Centre, Largo Galileo Galilei, 1, 00044 Frascati (RM), Italy}
\affiliation{INAF - Osservatorio Astronomico di Roma, Via Frascati, 33, 00040 Monte Porzio Catone (RM), Italy}

\author[0000-0002-2021-1863]{Jacqueline V. Keane}
\affiliation{Institute for Astronomy, 2680 Woodlawn Drive, Honolulu, HI 96822 USA}

\author[0000-0002-0439-9341]{Robert Weryk}
\affiliation{Institute for Astronomy, 2680 Woodlawn Drive, Honolulu, HI 96822 USA}

\author[0000-0002-1341-0952]{Richard Wainscoat}
\affiliation{Institute for Astronomy, 2680 Woodlawn Drive, Honolulu, HI 96822 USA}

\author[0000-0001-9701-4625]{Devendra K. Sahu}
\affiliation{Indian Institute of Astrophys., II Block, Koramangala, Bangalore 560 034, India}

\author[0000-0003-0174-3829]{Bhuwan C. Bhatt}
\affiliation{Indian Institute of Astrophys., II Block, Koramangala, Bangalore 560 034, India}

\begin{abstract}

2009 MS9 is a trans-Neptunian object (TNO) whose perihelion brings it close to the distance where some long period comets are seen to become active. Knowing this, and the fact that this object appears to brighten in excess of it's predicted nucleus brightness suggests that 2009 MS9 has a delayed onset of activity brought on by the sublimation of a species more volatile than water. In this paper we characterize 2009 MS9's physical properties and investigate potential outgassing through composite images, sublimation models, and measurements of spectral reflectivity. We find that deep composite images of the object at various epochs along its orbit show no evidence of dust yet place sensitive limits to the dust production. We estimate the nucleus radius to be 11.5 km $\pm 3.5$ km using thermal IR modeling from NEOWISE data and use this and data pre-perihelion to estimate a geometric albedo of 0.25. We compare a CO-sublimation activity model to its post perihelion heliocentric light curve and find this data supports an active fractional area of $5 \times 10^{-6}$ assuming 2 $\mu$m sized grains and other typical comet parameters. The spectral reflectivity of the surface materials obtained with the Gemini Observatory and CFHT at different epochs shows a reddening spectral slope. We compare the physical properties of 2009 MS9 to both TNO and comet populations, and speculate that 2009 MS9's reddening may be due to the buildup of a dust mantle on the surface and could be an explanation of why TNOs exhibit a color bimodality.

\end{abstract}
\keywords{minor planets, asteroids: individual (2009 MS9) --- comets: general}

\section{Introduction} \label{sec:intro}

The distribution and migration of volatiles in the solar system from past to present is a topic of great interest in astronomy because of its intimate connection to early solar system dynamics and chemistry and thus, the origin of water on earth. The most optimal way to investigate the solar system volatile distribution is to study the physics of primitive minor bodies, i.e. asteroids and comets. Primitive undifferentiated bodies are the least transformed volatile ``vehicles to the past''; because of their size, most lack the interior physical processes that obscure the chemical history imprinted since formation. Hence comets and asteroids can be used as tracers to infer what occurred during early solar system.

The volatiles that drive comet sublimation are H$_2$O, CO, and CO$_2$ ice \citep{meech2004}. These molecules produce outgassing that drags dust particles from the nucleus surface and sub-surface, often creating a visible coma which results in a change in brightness. This process becomes observable at different heliocentric distances characteristic of the volatile latent heat of sublimation, the effective surface temperature of the nucleus, and the dust grain sizes \citep{meech2004}. 

For example, the maximum rate of water-ice sublimation occurs at 180 K, but begins at lower rates with lower temperatures, hence ice sublimation can begin to lift micron-sized dust that can be detected as far as 5-6 au from the sun \citep{meech2004}. For CO$_2$, this occurs near 20 au with a peak sublimation rate at 80 K, and for CO at 25 K, this occurs beyond 50 au, at a distance beyond the Kuiper belt \citep{meech2004}. These values assume that the volatiles are located at the surface. However, because of thermal processing from previous perihelion passages or cosmic ray processing of comet surfaces prior to the comet entering the the inner solar system for the first time, volatiles are likely buried beneath the surface and do not begin sublimating until the heat wave from the surface penetrates the lower layers \citep{prialnik2004}.

\textcolor{black}{Because of this, volatiles likely exist in the comet as either differentiated layers (as a result of phase-transition, sublimation, and/or recondensation of ices) or pockets of crystalline or amorphous ice. In addition to traditional ice sublimation due to solar heating, out-gassing brought on by amorphous ice transition may is also possible for active bodies at large heliocentric distances. Pockets of amorphous ice are formed at a low pressure and temperature, from the process of rapid cooling or compression, such that that is no time to create a crystalline arrangement in the process. It is theorized that all comets have amorphous ice in their interior, as comets where likely formed in the conditions described above \citep{mekler1994}. Species more volatile than water can also exist in gaseous form, trapped this amorphous water-ice matrix, which has been shown by laboratory experiments to be an efficient way to trap large amounts of volatiles \citep{barnun1987}. When this amorphous ice crystallizes it releases heat and the trapped volatiles in the process. The heat triggers additional crystallization in surrounding areas leading to a run-away effect (the peak temperature for this being 120-130 K, or within 10 au), which could be a potential trigger for a sublimation outburst \citep{prialnik2004}.}

Until recently, scientists have not had a chance to thoroughly explore CO- or CO$_2$-driven activity, in part because previous surveys did not discover objects approaching perihelion far enough out to see them before they turned on, and in part because CO-rich comets are relatively rare \citep{meech2017}. Fortunately, over the past decade the University of Hawai`i has led the way in characterizing faint primitive body discoveries and follow-up observations using various telescopes and deep surveys.  These include the Canada-France-Hawaii Telescope (CFHT) Ecliptic Plane Survey (CFEPS, \citealt{jones2006}), the Pan-STARRS all-sky survey (PS1; \citealt{chambers2016}), the Asteroid Terrestrial-impact Last Alert System (ATLAS; \citealt{tonry2018}), and in the future the All-Sky Automated Survey for Supernovae (ASAS-SN) \citep{shappee2014} adapted for Near-Earth Objects (NEOs; Shappee, private communication). 

This explosion of deep all-sky surveys is currently creating a new avenue to probe solar system models with efficient detection and orbit characterization of faint objects. This gives us the opportunity to examine the volatiles over a much larger range of object types at a wider range of distances than before.  

2009 MS9 (asteroid 418993) was discovered using the Canada-France-Hawaii Telescope (CFHT) on 2009 June 25 as a part of the High Ecliptic Latitude (HiLat) extension of the Canada-France Ecliptic Plane Survey (CFEPS) \citep{petit2017}. It was found at a distance of 12.9 au from the Sun and an ecliptic latitude of 71$^{\circ}$ \citep{petit2017} and was classified as an exotic trans-Neptunian object (TNO) because of its highly eccentric ($e\sim$0.97) and inclined ($i\sim$68$^{\circ}$) orbit and its large semi-major axis ($a\sim$388 au) at epoch JD$=2459600.5$. Additionally, 2009 MS9 was identified in the Pan-STARRS1 survey of outer solar system objects as a distant solar object with one of the highest inclinations in the survey, with a maximum brightness $V_{max}$ $>$ 20.0, which is considered bright for a minor body with such large perihelion distance ($q$ = 10.99 au; \citealt{weryk2016}).

\citet{petit2017} obtained photometry of 2009 MS9 from 2009 to 2011. They conducted follow-up Palomar observations of the object in 2009 August and determined a single-peaked rotation period of $\sim$6.5 hours, or 13 hours double-peaked, with a maximum magnitude change of 0.4 mag \citep{petit2017}. Additional BRV colors were reported by \citet{jewitt2015} in their survey investigating colors of comets and related bodies. 

\begin{figure}[hb!]
\includegraphics[width=8cm]{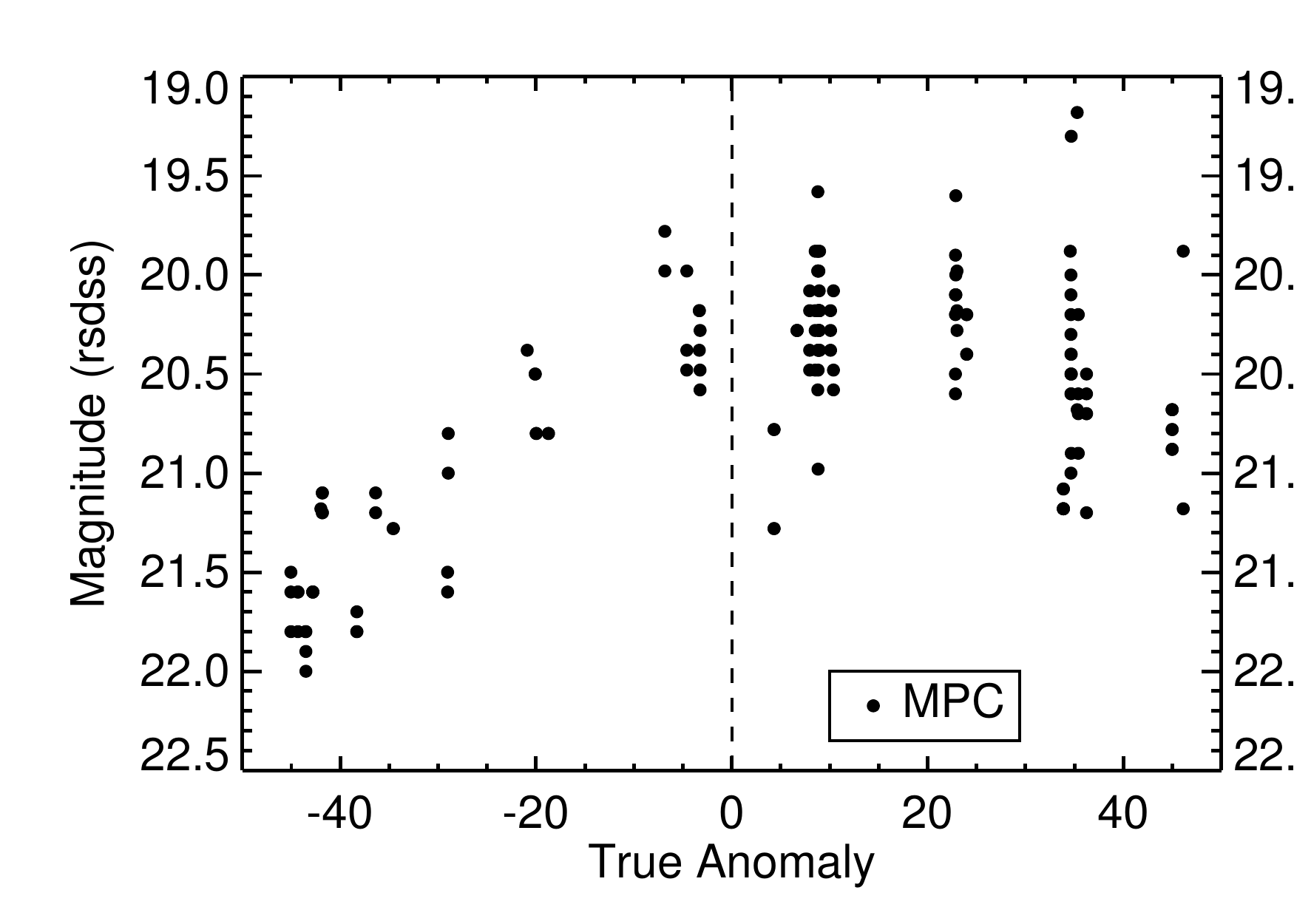}
\label{fig:heliompc}
\caption{The heliocentric light curve of 2009 MS9 shown with data from the Minor Planet Center (MPC), which consists primarily of photometry from \textcolor{black}{various sources, including} amateur telescopes. \textcolor{black}{The asymmetry of brightening post-perihelion is suggestive of activity} 
}
\end{figure}

Publicly available photometry of 2009 MS9 compiled through the Multitudinous Image-based Sky-survey and Accumulative Observations (MISAO) project \footnote{\url{http://www.aerith.net/misao/index.html}} \textcolor{black}{which gathers Minor Planet Center (MPC) data (reported values of amateur and professional observations from various observatories)} showed that \textcolor{black}{near the 2009 MS9's perihelion at Q=10.99 au on Feb 14 2013} the data did not seem to be following the brightness variations expected for an asteroid; that is, it experienced brightening post-perihelion, suggesting possible activity (see Fig. \ref{fig:heliompc}). Research has shown that many long period comets can become active around 10 au \citep{meech2009}; this is very similar to the perihelion of 2009 MS9 ($q\sim$10.99 au). Although officially classified as a TNO, because of this and its unique orbital properties, 2009 MS9 has been described as a Damocloid (\citealt{jewitt2015}; probable inactive long period comet) and a Manx candidate (object on a long period comet orbit with minimal activity and potential inner solar system origins; \citealt{meech2016}).

While there has been research investigating TNO activity \citep{meech2003}, and certainly such activity would be one explanation for the diversity of TNO and centaur colors \citep{hainaut2000, meech2003, sekiguchi2002}, there has been no direct evidence for sublimation to date \citep{young2020}, with the exception of the dwarf planet Pluto \citep{gladstone2019}. Hence if 2009 MS9 is found to have volatile activity, this would the first detection of its kind.

In this paper we attempt to verify volatile activity in 2009 MS9 using the following methods: 

\begin{itemize}
\item Using the heliocentric light curve photometry to compare this to a nucleus plus volatile activity light curve with an ice sublimation model;
\item Creating composite images to visually search for coma quantified with a comparison of 2009 MS9 and stellar radial profile fluxes; and
\item Analyzing colors pre- and post-perihelion to search for changes caused by activity-driven removal of surface materials.
\end{itemize}

We use an ensemble of photometric measurements taken from data obtained from several facilities, including multiple observatories on Maunakea. We describe the observations in detail in Section \ref{sec:observations} below. 

\section{Observations} 
\label{sec:observations}

\startlongtable
\begin{deluxetable*}{cccccccccc}
\tablecaption{Observing Geometry and Photometry \label{tab:data}}
\tablehead{
\colhead{UT Date                    } &
\colhead{JD                         } &
\colhead{$r$\tablenotemark{a}       } &
\colhead{$\Delta$\tablenotemark{a}  } &
\colhead{$\alpha$\tablenotemark{a}  } &
\colhead{TA\tablenotemark{b}        } &
\colhead{Filt                       } &
\colhead{\# Images                  } &
\colhead{avg mag (r)                    } &
\colhead{error                      }
}
\startdata
\multicolumn{8}{c}{\bf Gemini North/South} \\
\hline
2015-09-08 13:21:08 & 2457274.0563426 & 11.984 & 11.364 & 3.904 & 33.5 & griZ & 16 & 20.58 & 0.01 \\
2016-11-08 11:38:27 & 2457700.9850347 & 12.965 & 11.988 & 0.757 & 46.2 & r & 6 & 20.66 & 0.01\\
2018-01-17 07:23:18 & 2458135.8078472 & 14.187 & 13.958 & 3.894 & 57.1 & r & 4 & 21.50 & 0.01\\
2018-11-10 12:05:09 & 2458433.0035764 & 15.11 & 14.14 & 0.79 & 63.5 & griZY & 34 & 21.38 & 0.02\\
2018-11-30 06:54:40 & 2458452.7879630 & 15.17 & 14.28 & 1.67 & 63.9 & griZY & 35 & 21.48 & 0.02\\
2019-12-27 08:54:42 & 2458844.8713194 & 16.47 & 15.85 & 2.69 & 71.1 & r & 11 & 22.05 & 0.03\\
2020-02-20 05:58:42 & 2458899.7490972 & 16.66 & 16.89 & 3.28 & 72.0 & r & 7 & 22.17 & 0.04\\
2021-11-29 03:45:26 & 2459547.6565509 & 18.92 & 18.08 & 1.60 & 81.4 & griZ & 25 & 22.48 & 0.05 \\
\hline
\multicolumn{8}{c}{\bf CFHT } \\
\hline
2009-06-25 11:32:48 & 2455007.9811111 & 12.874 & 12.536 & 4.326 & -45.1 & r & 3 & 21.06 & 0.02\\
2009-07-19 08:33:28 & 2455031.8565741 & 12.814 & 12.456 & 4.316 & -44.4 & r & 4 & 21.12 & 0.02\\
2009-08-19 08:12:10 & 2455062.8417824 & 12.736 & 12.435 & 4.398 & -43.5 & r & 20 & 21.30 & 0.02\\
2009-08-21 06:37:33 & 2455064.7760764 & 12.731 & 12.436 & 4.406 & -43.4 & r & 33 & 21.30 & 0.03\\
2009-09-13 05:42:00 & 2455087.7375000 & 12.675 & 12.475 & 4.493 & -42.8 & r & 4 & 21.18 & 0.03\\
2009-10-12 05:06:51 & 2455116.7130903 & 12.605 & 12.565 & 4.539 & -42.0 & r & 5 & 21.13 & 0.03\\
2009-10-17 05:14:47 & 2455121.7185995 & 12.593 & 12.582 & 4.536 & -41.8 & r & 5 & 21.23 & 0.04\\
2009-11-14 05:47:30 & 2455149.7413194 & 12.526 & 12.681 & 4.442 & -41.0 & r & 9 & 21.16 & 0.04\\
2009-11-16 05:49:13 & 2455151.7425116 & 12.521 & 12.688 & 4.431 & -41.0 & r & 5 & 21.04 & 0.03\\
2010-02-13 15:52:31 & 2455241.1614699 & 12.316 & 12.709 & 4.151 & -38.3 & r & 3 & 21.20 & 0.06\\
2010-06-12 09:20:12 & 2455359.8890278 & 12.063 & 11.908 & 4.801 & -34.7 & r & 5 & 21.03 & 0.03\\
2010-07-07 08:15:23 & 2455384.8440162 & 12.012 & 11.745 & 4.734 & -33.9 & r & 8 & 20.91 & 0.02\\
2010-07-15 09:07:48 & 2455392.8804167 & 11.996 & 11.701 & 4.705 & -33.6 & gri & 108 & 21.10 & 0.03\\
2010-09-05 06:37:31 & 2455444.7760532 & 11.895 & 11.554 & 4.637 & -31.9 & r & 9 & 20.73 & 0.02\\
2010-09-11 07:14:28 & 2455450.8017130 & 11.884 & 11.554 & 4.650 & -31.8 & r & 25 & 20.80 & 0.02\\
2010-10-02 05:08:52 & 2455471.7144907 & 11.845 & 11.579 & 4.720 & -31.1 & gri & 27 & 20.97 & 0.03\\
2010-10-03 05:24:26 & 2455472.7253009 & 11.843 & 11.581 & 4.724 & -31.0 & gri & 27 & 20.91 & 0.03\\
2010-10-04 05:04:59 & 2455473.7117940 & 11.841 & 11.583 & 4.728 & -31.0 & gri & 27 & 20.82 & 0.03\\
2010-10-05 05:18:41 & 2455474.7213079 & 11.839 & 11.585 & 4.732 & -31.0 & gri & 28 & 21.06 & 0.03\\
2010-10-06 05:15:57 & 2455475.7194097 & 11.837 & 11.588 & 4.736 & -30.9 & gri & 27 & 20.76 & 0.02\\
2010-12-03 04:34:20 & 2455533.6905093 & 11.733 & 11.818 & 4.778 & -29.0 & r & 2 & 21.03 & 0.10\\
2011-07-31 12:32:14 & 2455774.0223843 & 11.369 & 11.062 & 4.946 & -20.8 & r & 45 & 20.92 & 0.03\\
2011-08-01 07:42:57 & 2455774.8214931 & 11.368 & 11.056 & 4.939 & -20.8 & r & 40 & 20.84 & 0.03\\
2011-08-02 11:12:35 & 2455775.9670718 & 11.367 & 11.049 & 4.930 & -20.7 & r & 45 & 20.85 & 0.03\\
2011-08-03 12:58:02 & 2455777.0403009 & 11.366 & 11.045 & 4.925 & -20.7 & r & 42 & 20.70 & 0.03\\
2013-07-04 14:06:33 & 2456478.0878819 & 11.027 & 11.159 & 5.210 & 5.32 & r & 3 & 20.67 & 0.03\\
2013-07-05 13:22:40 & 2456479.0574074 & 11.027 & 11.147 & 5.220 & 5.36 & r & 1 & 20.51 & 0.01\\
2013-07-06 13:34:43 & 2456480.0657755 & 11.027 & 11.136 & 5.230 & 5.40 & r & 1 & 20.56 & 0.01 \\
2014-09-26 14:18:27 & 2456927.0961458 & 11.406 & 10.614 & 3.208 & 21.8 & r & 5 & 20.25 & 0.01\\
2016-02-06 05:09:16 & 2457424.7147685 & 12.300 & 12.527 & 4.431 & 38.2 & r & 2 & 20.94 & 0.05\\
2016-09-29 14:55:21 & 2457661.1217708 & 12.861 & 12.026 & 2.556 & 45.0 & r & 3 & 20.86 & 0.01\\
2017-08-21 14:38:26 & 2457987.1100231 & 13.747 & 13.489 & 4.118 & 53.6 & r & 3 & 21.26 & 0.02\\
2017-09-13 14:52:18 & 2458010.1196528 & 13.813 & 13.205 & 3.405 & 54.1 & r & 2 & 21.06 & 0.02\\
2017-12-12 09:18:40 & 2458099.8879630 & 14.078 & 13.317 & 2.608 & 56.2 & r & 4 & 21.13 & 0.01\\
2018-02-13 05:51:12 & 2458162.7438889 & 14.268 & 14.494 & 3.830 & 57.7 & r & 2 & 21.51 & 0.03\\
2018-11-05 10:58:27 & 2458427.9572569 & 15.095 & 14.124 & 0.796 & 63.4 & r & 3 & 21.33 & 0.03\\
2020-09-11 14:12:22 & 2459104.0919213 & 17.364 & 16.938 & 3.049 & 75.2 & r & 3 & 22.45 & 0.06\\
2020-12-17 08:26:20 & 2459200.8516204 & 17.702 & 16.974 & 2.185 & 76.7 & r & 3 & 22.45 & 0.03\\
2021-01-12 05:56:41 & 2459226.7476968 & 17.793 & 17.383 & 2.910 & 77.0 & r & 3 & 22.60 & 0.03\\
2021-02-13 05:30:00 & 2459258.7291667 & 17.905 & 17.989 & 3.140 & 77.5 & r & 2 & 22.55 & 0.04\\
2021-08-05 14:38:23 & 2459432.1099884 & 18.514 & 18.687 & 3.079 & 79.9 & r & 2 & 22.52 & 0.03\\
2021-10-08 12:34:24 & 2459496.0238889 & 18.741 & 18.048 & 2.248 & 80.7 & r & 4 & 22.66 & 0.05\\
2021-10-10 12:22:54 & 2459498.0159028 & 18.748 & 18.037 & 2.189 & 80.7 & r & 4 & 22.86 & 0.03\\
2022-02-27 05:48:28 & 2459637.7416666 & 19.246 & 19.500 & 2.832 & 82.5 & r & 2 & 22.76 & 0.07 \\
2022-08-25 14:45:08 & 2459817.1145833 & 19.886 & 19.790 & 2.907 & 84.6 & r & 2 & 22.83 & 0.05\\
\hline
\multicolumn{8}{c}{\bf Pan-STARRS \tablenotemark{c}} \\
\hline
2011-08-15 10:03:03.9 & 2455788.9187958 & 11.351 & 10.972 & 4.824 & -20.3 & i & 2 & 20.74 & 0.06\\
2011-08-20 08:51:11.4 & 2455793.8688821 & 11.345 & 10.949 & 4.789 & -20.1 & ri & 4 & 20.61 & 0.03\\
2011-08-23 07:55:56.1 & 2455796.8305093 & 11.341 & 10.935 & 4.766 & -20.0 & gr & 4 & 20.84 & 0.06\\
2011-09-27 06:38:25.1 & 2455831.7766782 & 11.300 & 10.851 & 4.638 & -18.7 & r & 2 & 20.61 & 0.04\\
2012-08-16 11:57:00.4 & 2456155.9979167 & 11.042 & 10.646 & 4.929 & -6.81 & gr & 4 & 20.55 & 0.04\\
2014-08-13 14:38:49.6 & 2456883.1103009 & 11.349 & 11.042 & 4.945 & 20.3 & i & 4 & 20.73 & 0.03\\
2014-09-29 12:09:05.0 & 2456930.0063079 & 11.410 & 10.599 & 3.071 & 21.9 & w & 3 & 20.59 & 0.02\\
2014-11-27 05:56:05.2 & 2456988.7472801 & 11.491 & 10.783 & 3.534 & 24.0 & r & 1 & 20.55 & 0.03\\
2014-11-29 05:59:31.7 & 2456990.7496759 & 11.495 & 10.806 & 3.628 & 24.1 & i & 4 & 20.58 & 0.06\\
2014-12-11 08:25:38.0 & 2457002.8511343 & 11.512 & 10.959 & 4.153 & 24.5 & r & 3 & 20.62 & 0.03\\
2015-08-30 12:03:22.7 & 2457265.0023495 & 11.966 & 11.467 & 4.297 & 33.2 & i & 2 & 20.81 & 0.10\\
2015-09-10 15:09:59.8 & 2457276.1319444 & 11.988 & 11.342 & 3.803 & 33.6 & z & 1 & 20.90 & 0.10 \\
2015-11-03 12:12:14.4 & 2457330.0084954 & 12.096 & 11.147 & 1.412 & 35.3 & r & 4 & 20.56 & 0.06\\
2015-12-05 06:02:23.2 & 2457361.7516551 & 12.163 & 11.427 & 3.179 & 36.3 & r & 4 & 20.58 & 0.02\\
2015-12-06 05:55:54.4 & 2457362.7471528 & 12.165 & 11.440 & 3.236 & 36.3 & w & 4 & 20.72 & 0.02\\
2016-01-22 06:24:06.3 & 2457409.7667361 & 12.266 & 12.243 & 4.600 & 37.8 & i & 3 & 20.78 & 0.07\\
2016-07-17 14:24:34.6 & 2457587.1004051 & 12.678 & 12.924 & 4.414 & 43.0 & w & 4 & 21.16 & 0.03\\
2017-01-23 06:31:58.2 & 2457776.7721991 & 13.164 & 13.100 & 4.285 & 48.3 & w & 4 & 21.29 & 0.03\\
2017-08-03 14:37:30.4 & 2457969.1093750 & 13.697 & 13.729 & 4.238 & 53.2 & w & 4 & 21.47 & 0.04\\
2017-09-23 11:19:36.1 & 2458019.9719444 & 13.844 & 13.104 & 2.891 & 54.4 & w & 4 & 21.31 & 0.03\\
2017-10-19 13:49:29.6 & 2458046.0760417 & 13.920 & 12.965 & 1.204 & 55.0 & w & 4 & 20.82 & 0.06\\
2017-11-07 11:27:42.1 & 2458064.9775694 & 13.976 & 12.989 & 0.378 & 55.5 & i & 4 & 20.82 & 0.04\\
2017-11-16 11:10:40.4 & 2458073.9657407 & 14.003 & 13.040 & 0.962 & 55.7 & w & 2 & 21.19 & 0.07\\
2017-11-21 11:06:09.0 & 2458078.9626042 & 14.017 & 13.080 & 1.311 & 55.8 & w & 4 & 21.18 & 0.02\\
2018-01-12 07:46:36.2 & 2458130.8240278 & 14.172 & 13.859 & 3.810 & 57.0 & w & 4 & 21.50 & 0.04\\
2018-01-30 06:22:10.6 & 2458148.7654051 & 14.226 & 14.217 & 3.968 & 57.4 & i & 4 & 21.57 & 0.10\\
2019-01-05 07:28:55.2 & 2458488.8117477 & 15.293 & 14.824 & 3.287 & 64.6 & w & 4 & 22.06 & 0.07 \\
2019-10-24 12:32:44.2 & 2458781.0227315 & 16.260 & 15.356 & 1.510 & 70.0 & w & 4 & 21.85 & 0.05 \\
2019-12-03 08:44:49.8 & 2458820.8644676 & 16.395 & 15.538 & 1.747 & 70.7 & i & 4 & 22.02 & 0.09 \\
2019-12-20 09:06:04.7 & 2458837.8792245 & 16.452 & 15.752 & 2.455 & 71.0 & w & 4 & 22.09 & 0.06 \\
\hline
\multicolumn{8}{c}{\bf HCT } \\
\hline
2015-12-04 17:53:20.00 & 2457361.2453704 & 12.160 & 11.411 & 3.110 & 36.2 & R & 3 & 20.78 & 0.03\\
2015-12-05 14:43:25.00 & 2457362.1134838 & 12.162 & 11.424 & 3.168 & 36.3 & R & 13 & 20.67 & 0.02\\
2016-12-03 16:57:13.00 & 2457726.2064005 & 13.028 & 12.210 & 2.494 & 46.9 & R & 9 & 21.00 & 0.03\\
\hline
\multicolumn{8}{c}{\bf Lowell } \\
\hline
2013-10-04 06:46:18.01 & 2456569.7821529 & 11.067 & 10.320 & 3.566 & 8.79 & r & 10 & 20.22 & 0.01\\
\hline
\enddata
\tablenotetext{a}{Heliocentric, geocentric distance [au]; and phase angle [deg]}
\tablenotetext{b}{True anomaly [deg], the position along orbit; TA at perihelion = 0$^{\circ}$}
\tablenotetext{c}{filters reported for PS1 data are the PS1 g,r,i,z filters}
\tablenotetext{d}{Magnitude and error through 5$''$ radius aperture; and converted to SDSS r$'$ as described in the text}
\end{deluxetable*}

The geometry of all the observations is detailed in Table \ref{tab:data}. We processed all of the ground-based CCD imaging data using the same technique and tools as described in \citet{meech2017}: we use the Terapix/Astromatic tools \citep{bertin1996} to fit world coordinates (RA and Dec) based on reference stars from the 2MASS catalog. We used SExtractor \citep{bertin1996} automatic apertures to measure the magnitudes of trailed stars and computed a photometric zero point for each image based on stars from the PS1 database \citep{flewelling2020,magnier2020b} 3$\pi$ Steradian Survey \citep{chambers2016}. Lastly, we used Terapix tool (SExtractor) to produce multi-aperture target photometry. We visually inspected each image and removed those contaminated by background objects, bad pixels, or cosmic rays. We used a 4$\arcsec$ aperture for photometry, as that yielded the most uncontaminated images for analysis as many of our observations where in crowded fields. 

We convert all of our magnitudes to the sdss r filter shown in Table \ref{tab:data} using the following filter transformations:

From $R$ Cousins system magnitudes to r, we used the transformations of \citet{lupton2005}
\begin{equation}
    \begin{split}
    r = R + 0.1837(g - r) + 0.0971 \\ \sigma_{r} = \sigma_{R}^2 + (0.1837\sigma_{(g-r)})^2 + 0.0106^2
    \end{split}
\end{equation}

The transformations from the PS1 filters to r are given by \citet{tonry2012}.  
\begin{equation}
    \begin{split}
	r = g_{P1} + 0.011 - 0.875(g-r) + 0.015 \\ {(g-r)}^2,  
	\sigma_{r} = (0.006^2 + \sigma_{gp1}^2)^{0.5}
	\end{split}
\end{equation}
\begin{equation}
    \begin{split}
	r = r_{P1} - 0.001 + 0.006(g-r) + 0.002 {(g-r)}^2, 
	\\ \sigma_{r} = (0.002^2 + \sigma_{rp1}^2)^{0.5} 
    \end{split}
\end{equation}
\begin{equation}
    \begin{split}
	r = i_{P1} + (r-i) - 0.004 + 0.014(g-r) - \\ 0.001 {(g-r)}^2,
	\sigma_{r} = (0.003^2 + \sigma_{ip1}^2)^{0.5}
    \end{split}
\end{equation}
\begin{equation}
    \begin{split}
	r = z_{P1} + (r-z) + 0.013 - 0.040(g-r) + \\ 0.001 {(g-r)}^2 
	\sigma_{r} = (0.009^2 + \sigma_{zp1}^2)^{0.5}
    \end{split}
\end{equation}
\begin{equation}
    \begin{split}
	r = w_{P1} - 0.018 - 0.118(g-r) + 0.091 {(g-r)}^2 
	\\ \sigma_{r} = (0.012^2 + \sigma_{wp1}^2)^{0.5}
    \end{split}
\end{equation}

We report on the average magnitude for nights where multiple images in the same filter are taken, and these magnitudes are used to construct the target's heliocentric light curve (see section \ref{sec:sublimation}). \textcolor{black}{Color corrections for each R observation are made based off of the target's gemini or cfht colors (e.g. g-r) closest to when the data was taken.}

\subsection{Canada-France-Hawaii Telescope (CFHT)}

Data was obtained from the Canada-France-Hawaii MegaCam wide-field imager, an array of forty 2048$\times$4612 pixel CCDs with a plate scale of 0$\farcs$187 per pixel and a 1.1 square degree FOV. \citet{petit2017} obtained 636 observations of 2009 MS9 over the 40 nights between the dates of 2009 June 25 and 2016 February 6 as follow up photometry to the 2009 June discovery. Additionally in 2010 July and October several nights of $gri$ magnitudes where obtained pre-perihelion \citep{petit2017}. Our team also obtained photometry with MegaCam during 2014, 2017, 2018, 2020, and 2021.

\subsection{Gemini Observatory}

Photometric observations of 2009 MS9 were taken on six nights between 2015 and 2021, and $griz$ or $griZY$ colors were obtained on four of those dates. Data were obtained from the Gemini North 8m telescope on Maunakea or the Gemini South 8m telesope with the Gemini Multi-Object Spectrograph (GMOS) \citep{hook2004} in imaging mode. GMOS is a mosaic of three 2048$\times$4176 Hamamatsu detectors, binned 2$\times$2. The data were obtained through SDSS filters using queue service observing and were processed to remove the instrumental signature using DRAGONS, Gemini Observatory's Python-based data reduction software \citep{labrie2019}. For nights where the outer chips were non-photometric due to guide probe vignetting, we extracted only the central chip of the mosaic so that the zero-point calibration would not be affected.

\subsection{Lowell 1.8m Perkins Telescope}

Data were obtained during 2013 October 4 and 5 using the 1.8m Perkins telescope at Lowell Observatory with the Perkins ReImaging SysteM (PRISM) in imaging mode. The 2$\times$K E2V CCD has 0$\farcs$39 pixels, with a gain of 2.72 e-/ADU and read noise 7.2 e-.  The data were guided at non-sidereal rates under photometric conditions, but with mediocre seeing (2.0$\arcsec$ FWHM) on both nights.

\subsection{Himalayan Chandra Telescope (HCT)}

As part of our Manx characterization efforts, we have a long-term program on the 2.01 m HCT at Mt. Saraswati, Hanle, India to obtain images for astrometry and heliocentric light curves. We used the Himalaya Faint Object Spectrograph and Camera (HFOSC) with the Bessell/Cousins filter system to obtain data on the two dates shown in Table \ref{tab:data}. The detector has a read noise of 4.8 e- and gain of 1.22 e-/ADU and was read out in 1$\times$1 binning mode with a pixel scale of 0$\farcs$296/pixel. This telescope cannot autoguide at non-sidereal rates, so we kept exposures short enough to keep trailing to less than the typical seeing while guiding at sidereal rates. 

\subsection{NEOWISE}

The NEOWISE mission uses the reactivated WISE spacecraft, which began survey operations in 2013 December \citep{mainzer2014}. Without any cryogens, only the two short-wavelength channels at 3.4~$\micron$ (W1) and 4.6~$\micron$ (W2) are available. Thirty four exposures of 2009 MS9 were obtained between 2010 May 21-23 (mid-obs time 2010-05-22T20:07:27.597) for a total exposure time of 300 s.

\subsection{Pan-STARRS1}

Pan-STARRS1 astrometric positions of 2009 MS9 were identified by the Moving Object Processing Pipeline \citep{denneau2013} and photometry was extracted using the Pan-STARRS Image Processing Pipeline \citep{magnier2020b}. Pan-STARRS photometry was provided in $g_{P1}$,$r_{P1}$ and $i_{P1}$ Pan-STARRS filters, by Point-Spread-Function photometry in ubercalibrated chip-stage images \citep{schlafly2012}, with associated measurement uncertainties.

\section{Data Analysis}  \label{sec:reduction}

\subsection{WISE Nucleus Size Modeling}
\label{sec:WISE}
The WISE data were processed through the WISE science data pipeline \citep{wright2010} to bias-subtract, flatten the images, and remove artifacts. The 34 images are then stacked using the comet's apparent rate of motion. In order to produce a spectral energy distribution, aperture photometry is converted to fluxes using the WISE zero-points and appropriate color temperature corrections \citep{wright2010}. These corrections are temperature dependent, and an initial guess at the temperature is required based on the expected blackbody temperature for the heliocentric distance of the observation. The data were acquired during the warm mission, and we use a technique that has been modified by \citet{reach2013} for the Spitzer warm mission to work with only the two shortest wavelength bands, W1 and W2. \textcolor{black}{We measure the flux of 2009 MS9 from the image, then the size of the target is measured with this flux in combination with a thermal model.}

\textcolor{black}{Initially \cite{Lebofsky1986} developed the Standard Thermal Model (STM) to determine the thermal flux of inactive minor bodies. This model uses a target's orbital properties and optical/thermal phase coefficients and is quantified by a ``beaming parameter.'' The beaming parameter accounts for the difference between isotropic and non-isotropic thermal emission of minor bodies in the infrared. The Near Earth Asteroid Thermal Model (NEATM) arose from allowing the thermal phase coefficient of the target to vary as a free parameter that can be fit when two or more infrared bands are available \citep{harris1998}.} \textcolor{black}{ \cite{Wright2007} has found that in comparison to a more sophisticated thermal model the NEATM model and associated beaming parameter gives diameters that are accurate to within 10\%. For Jupiter-family comets and outer solar system objects the average beaming parameter was measured to be $1.03 \pm 0.11$ and $1.2 \pm 0.35$ \citep{fernandez2013, stansberry2008}.} \textcolor{black}{We use the NEATM thermal model and determine the beaming parameter using a least-squares minimization in comparison to the WISE detections \citep{mainzer2011}.} 
This yields a beaming parameter of 0.95$\pm$0.3 \citep{bauer2015} and  a resulting radius of $R_N$ = 11.5$\pm$3.5 km. \textcolor{black}{Using this radius, we can combine this with the MPC's absolute magnitude of $H=9.8$, and use the traditional relationship between absolute magnitude (H), nucleus diameter (D), and geometric albedo (p) to recover $p$:}

\begin{equation}
\label{eq:hvalue}
D = \frac{1329}{\sqrt{p}}10^{-0.2H}
\end{equation}

This yields an estimated albedo of $p$ = 0.40$\pm$0.14. \textcolor{black}{However, because the H values reported by the MPC are not always accurate, we independently derive the albedo for 2009 MS9 in section \ref{sec:albedo}}

\subsection{Composite Images}
\label{sec:comp}

\begin{figure*}
\begin{center}
\includegraphics[width=17cm]{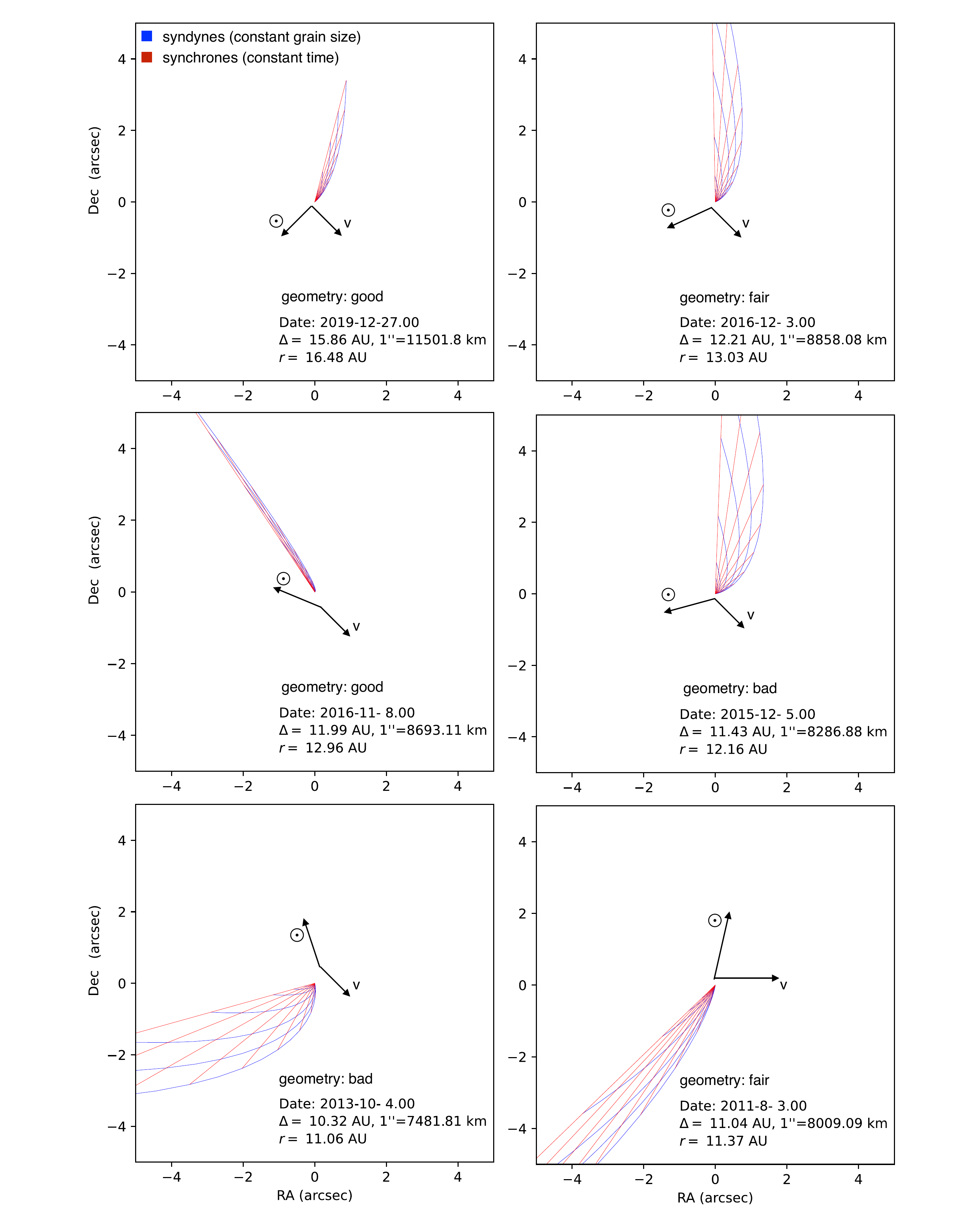}
\caption{FP modeling of the dust for 2009 MS9 for different epochs. \textcolor{black}{Each panel is labeled with the viewing geometry quality (bad, fair, good). Bad quality represents when the viewing geometry was far from the plane of the orbit. Good quality represents when the viewing geometry is close to the orbital plane. For the latter dust is more likely to be detected because the synchrones and syndynes are concentrated and overlap in a smaller area. }
}
\label{fig:FP}
\end{center}
\end{figure*}

Deep composite images are powerful tools for detecting weak activity that may not be apparent in individual images. This is especially important for characterizing activity in comets where a dust tail is not immediately apparent. We thus create composite images of 2009 MS9 at several epochs in an attempt to detect activity and/or place limits on activity in its orbit in the event none is detected. \textcolor{black}{Because this object has data that spans over a decade, we use Finson-Probstein plots, exposure time, our object's orbital position, and telescope capabilities to evaluate which observations are the most likely to yield dust from composite images.} 

\begin{table*}[ht!]
\caption{\label{table:dust}Composite Images and Dust Limits}
\begin{center}
\begin{tabular}{|lllcccc|ccc|ccc|}
%\hline
\hline
\# & Date & Obs. & Exp. & Dust & TA & r &\multicolumn{3}{c|}{Dust Limit (kg/s)} & \multicolumn{3}{c|}{Predicted Dust (kg/s)} \\
%\hline
 & & &(tot) & View &(deg)&(au)&0.5 $\micron$ & 2 $\micron$ & 10 $\micron$ & 0.5 $\micron$ & 2 $\micron$ & 10 $\micron$\\
\hline
%\hline
1 & 8/19-21 2009 & CFHT & 6500 & fair & -44 & 12.74 & $0.022_{-0.007}^{+0.007}$ & $0.088_{-0.027}^{+0.028}$ & $0.441_{-0.133}^{+0.138}$ & 0.015 & 0.059 & 0.355 \\ 
2 & 9/11 2010 & CFHT & 6250 & bad & -32 & 11.88 & $0.021_{-0.006}^{+0.009}$ & $0.085_{-0.025}^{+0.035}$ & $0.425_{-0.124}^{+0.176}$ & 0.018 & 0.069 & 0.411 \\
3 & 7/31-8/1 2011 & CFHT & 6000 & fair & -21 & 11.37 & $0.040_{-0.016}^{+0.042}$ & $0.159_{-0.063}^{+0.169}$ & $0.794_{-0.316}^{+0.730}$ & 0.020 & 0.075 & 0.450\\ 

4 & 8/2-3 2011 & CFHT & 6000 & fair & -21 & 11.37 & $0.031_{-0.011}^{+0.042}$ & $0.123_{-0.045}^{+0.080}$ & $0.616_{-0.223}^{+0.402}$ & 0.020 & 0.075 & 0.450\\ 
\hline
\hline
5 & 10/4 2013 & Lowell & 8100 & bad & 9 & 11.07 & $0.009_{-0.003}^{+0.002}$ & $0.035_{-0.010}^{+0.006}$ & $0.177_{-0.052}^{+0.030}$ & 0.021 & 0.079 & 0.477\\ 
6 & 9/8 2015 & Gemini & 800 & fair & 33 & 11.98 & $0.032_{-0.014}^{+0.027}$ & $0.127_{-0.056}^{+0.106}$ & $0.633_{-0.280}^{+0.532}$ & 0.018 & 0.068 & 0.408\\ 
7 & 12/4-5 2015 & HCT & 4500 & bad & 36 & 12.16 & $0.028_{-0.007}^{+0.012}$ & $0.111_{-0.026}^{+0.047}$ & $0.554_{-0.130}^{+0.233}$ & 0.017 & 0.066 & 0.395\\ 
8 & 11/8 2016 & Gemini & 900 & good & 46 & 12.96 & $0.014_{-0.005}^{+0.011}$ & $0.057_{-0.021}^{+0.044}$ & $0.283_{-0.104}^{+0.220}$ & 0.015 & 0.058 & 0.347\\ 
9 & 12/3 2016 & HCT & 2700 & fair & 46 & 13.03 & $0.054_{-0.019}^{+0.048}$ & $0.218_{-0.077}^{+0.048}$ & $1.09_{-0.384}^{+0.239}$ & 0.015 & 0.057 & 0.347\\ 
10 & 12/27 2019 & Gemini & 3000 & good & 71 & 16.48 & $0.021_{-0.010}^{+0.009}$ & $0.084_{-0.040}^{+0.034}$ & $0.421_{-0.200}^{+0.172}$ & 0.009 & 0.035 & 0.212\\ 
11 & 2/20 2020 & Gemini & 1000 & good & 72 & 16.66 & $0.025_{-0.009}^{+0.077}$ & $0.099_{-0.034}^{+0.306}$ & $0.494_{-0.170}^{+1.53}$ & 0.009 & 0.035 & 0.207\\ 
\hline
\end{tabular}
\end{center}
\vspace{0.2cm}
{\bf Notes:} Dates and observing circumstances for which we have created composite images of 2009 MS9. \textcolor{black}{The table is split into pre- and post-perihelion observations seperated by double horizontal lines.}  ``Dust view'' refers to whether or not the object is observed with a predicted dust tail close to the plane (see Figure \ref{fig:FP} for reference). The dates that yield the most favorable circumstances for detecting dust are denoted as good, with less favorable circumstances denoted as fair and bad. Numbers after each date correspond to radial profiles shown in Figure \ref{fig:dust} (see text.) The last two columns show the dust limits at the background limit from each composite's azimuthally averaged profiles in comparison to the dust production rate from sublimation modeling assuming a 1-1 gas-to-dust ratio.
\end{table*}

\textcolor{black}{Finson-Probsten (FP) plots \citep{finson1968} represent the trajectories of dust emitted from the nucleus of an active body at a specific time of it's orbit, governed by the forces of solar radiation pressure and gravity. They consist of synchrones (lines of constant time) and syndynes (lines of constant grain sizes) to map out a target's expected dust tail map, assuming the emission velocities of the dust are negligable. Thus an FP plot, while not exact, is a good approximation of the expected behavior of the tail for a given position in a comet's orbit.}

\textcolor{black}{We can use FP plots to our advantage to strategically pick out observations where the signal to noise ratio of the dust tail will be maximized. A non-favorable viewing geometry is when the dust is spread out over a larger region of the sky, and thus the surface brightness of the comet tail is lower (i.e. \ref{fig:FP}, top panel). The detection of dust is most likely when an active body has a tail that is edge on to the observer (i.e. \ref{fig:FP}, lower panel). This is when the dust is concentrated in the plan of the dust emission (the comet's orbital plane), thus the signal to noise ratio is high because the synchrones and syndynes will gather over a smaller area of the sky. }\textcolor{black}{Keeping this in mind, we generate FP diagrams to map out the positions of dust grains for active bodies in the solar system.} 

The deeper the composite (e.g. the longer the effective exposure time), the more likely we are able to detect dust from activity. Hence we also choose observations taken with large aperture telescopes on nights with the most images, giving the longest exposures. \textcolor{black}{Next, We selected observations close to perihelion, where the input of energy is the greatest which would translate to a larger gas/dust production rate, and post-perihelion, where 2009 MS9 is most likely to become active following a delayed outburst of activity (see section \ref{sec:albedo} for further discussion of this).} Our inspection of 2009 MS9's photometric data shows that the largest deviation from the fit nucleus light curve occurs most prominently post-perihelion, after TA $\sim$ 20 (see Fig.~\ref{fig:sublimation}). 

The list of observations we have created composite images for, as well as their observing circumstances and the target's orbital properties, are stated in Table~\ref{table:dust}, with the dust viewing conditions quantified as good, fair, or bad by the predicted level of concentration of dust near the orbital plane \textcolor{black}{as estimated by the FP plots.} \textcolor{black}{Columns seven and eight from Table~\ref{table:dust} result from this paper's modeling and are discussed in further detail in section \ref{sec:dustmodeling}}

\begin{figure}[ht!]
\includegraphics[width=8.35cm]{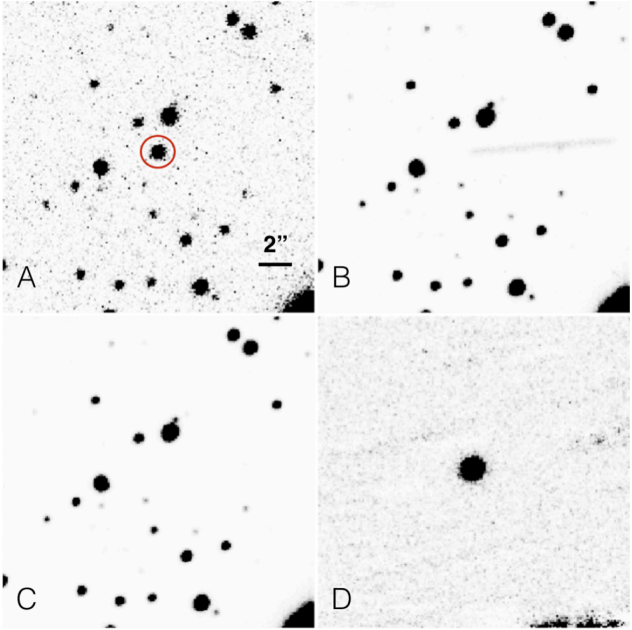}
\vspace{-.5cm}
    \caption{The process of creating a composite r-band image of 2009 MS9 using 45 images from CFHT obtained on July 31st 2011. A: one frame with an exposure time of 120 seconds. 2009 MS9 is highlighted by red circle. B: a sum composite image where 2009 MS9 is seen as trailed. C: the star template from an median composite image for the night of 2011-07-31 which is combined with 2011-08-01 used for stellar radial profiles and subtraction. D: The average composite image shifted by the KBO apparent rate of motion or 2009 MS9's astrometric position per night.}
\label{fig:compprocess}
\end{figure}

The creation of composite images requires several steps of data reduction to accurately prepare the images in an attempt to detect comet dust in a faint target. \textcolor{black}{First, we flag and remove single images where the 2009 MS9 overlaps a background object, bad pixel, or cosmic ray.} Next, we ``super-flatten'' each image using SExtractor to approximate and subtract the shape and amplitude of the background for each observation such that the background is approximately zero. Accurate subtraction of the background is critical. Because of this we iterate through different values of SExtractor's background parameters ``BACK\_SIZE'' and ``BACK\_FILTER'' until the flatness of the background is optimized. If BACK\_SIZE and/or \_FILTER have values that are too small, this will interfere with the wings of bright stars and risks subtraction of potential coma surrounding 2009 MS9. Because of this we take care to manually inspect the values of BACK\_SIZE and \_FILTER for each composite image, and the value of BACK\_SIZE does not go under 84 pixels in diameter. We then apply a bad-pixel mask to each image to get rid of persistent artifacts, bad pixels or cosmic rays. For each night we align the images to PS1 reference stars used for calibration to compensate for dithering. An example of a single reduced image is shown in (Fig. \ref{fig:compprocess}A).

We median combine the images to create a star template for subtraction (Fig. \ref{fig:compprocess}B). As the PSF of MS9 is often larger than the apparent velocity of the object and time elapsed between observations, a relic of 2009 MS9 is still prevalent in star templates. We alleviate this issue by manually replacing the minor body with simulated background noise using routines in IRAF for several images in one night's stack (Fig. \ref{fig:compprocess}C). We then subtract the star template from each image and then shift each image according to 2009 MS9's apparent rate of motion or according to its astrometric position. 

We median combine the 2009 MS9-shifted images to create the final composite stack with just 2009 MS9 in the observation (Fig. \ref{fig:compprocess}D). If there are several consecutive nights, we execute the above process for each individual night, and lastly align each night's 2009 MS9 composite with astrometrically calculated positions to create the final stack. 

Nearly half of our composites were created manually, while the rest were created using a automated version of the above methods, with the use of SWARP \citep{swarp2010} for the image shifts and stacking, and an automated Sextractor background subtraction built into the routine. We create 11 stacks total at different epochs or positions along 2009 MS9's orbit. We do not visually detect dust in any of our composite images. However, this can be quantified with an analysis comparing the flux profile of 2009 MS9 with those of field stars, and this is used to calculate a limit on the dust production, as is discussed in detail below.

\subsection{Dust Modeling}
\label{sec:dustmodeling}

While a deep composite image can provide visual confirmation of a dust coma, this property can be qualified by inspection of 2009 MS9's radial brightness profile in comparison to the average stellar profile of stars created from a star template. The deviation of 2009 MS9 from stellar profile will yield sensitive limits on the amount of dust at that given epoch. 

A comet with no detectable coma will have a near-identical azimuthally-averaged radial profile to stars in the same field, with an associated error. This error can be used to determine the upper-limit on dust production, or the maximum flux contribution from scattered coma light \citep{meech2003}. We use 3-$\sigma$ to define this upper limit. This flux, $F$, is related to the dust production rate $Q_{p}$ s$^{-1}$ by: 

\begin{equation}
    F=S_{\odot}\pi a^{2}_{gr}p_{v}Q_{p}\phi / 2r^{2}\Delta^{2}v_{gr}
\end{equation}

\noindent where $S_{\odot}$ is the solar flux in the appropriate filter, $a_{gr}$ the average grain size, $p_v$ the grain albedo, $\phi$ the projected diameter of the aperture [m], $r$ and $\Delta$ are the heliocentric and geocentric distances, and  $v_{gr}$ is the grain velocity. The total dust production $Q$ is then $Q_{p} \times $ grain mass.

\textcolor{black}{We derive the flux from our surface brightness profiles. Using this we can use equation 8 to derive the dust production rate assuming a volatile species such as CO or CO2. We can use this dust production rate to set limits on our sublimation modeling (see Section \ref{sec:sublimation})}. 

We construct the radial profiles of our target by measuring the target's flux in increasingly large apertures or annuli; the radius of each annulus increases by one pixel out to 30 pixels from the center which encapsulates the target plus some background. We convert this to magnitudes per arcsec$^2$ after photometrically calibrating the target image using the zero-point from the associated star template combined with the plate scale (arcsec/pixel) unique to the telescope used for each composite image. 

We then implement the same technique for two to five stars of comparable brightness in the star templates associated with each composite image, and used this to create an average flux profile for the stars. We then compare this to the relative flux profile of 2009 MS9 to see if there is qualitative evidence of a coma. 

To assure that both our manual and automated methods produce consistent results, we implement both for our 2010 September and 2015 September composite images and compare. We find that both techniques yield radial profiles that are nearly identical, and yield dust limits that deviate less than 0.05 kg s$^{-1}$ from one another, or deviate by $<$ 5\% around 1$\arcsec$ from the center of the profile. Hence we are confident our automated method and manual method of creating composites are consistent. 

\begin{figure*}
\begin{center}
\includegraphics[width=18cm]{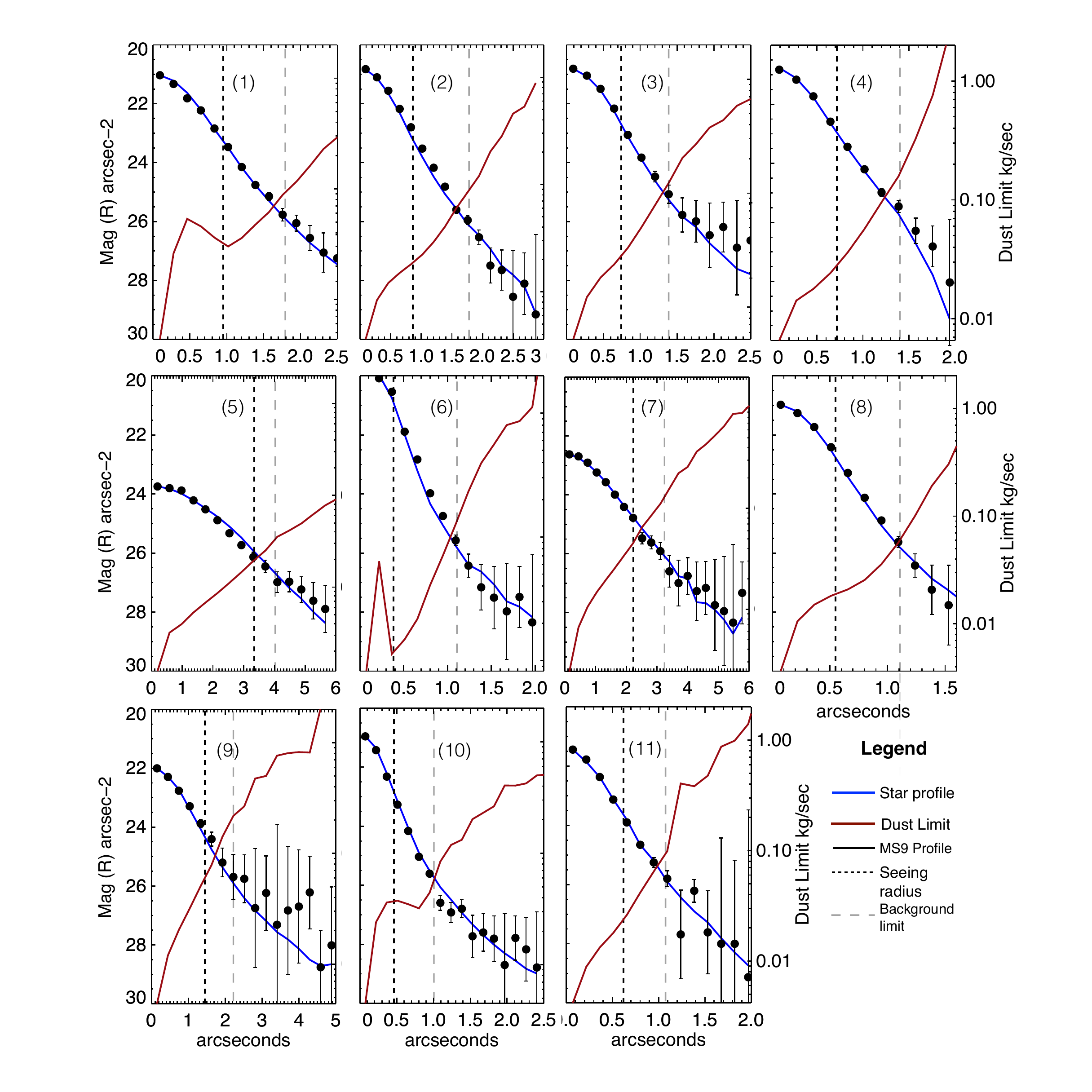}
\caption{For each composite image we bin the photon flux as a function of radius and calculate the average stellar radial profile of 2009 MS9 (black) vs the average star profile (blue). From these profiles we can calculate the limits on dust production (red).  
}
\label{fig:dust}
\end{center}
\end{figure*}

Each panel in Fig. \ref{fig:dust} represents the radial profile results for each composite image (11 total). This includes 2009 MS9's radial profile (black) in comparison to the average stellar profile (blue) and the resulting limit on dust production (red) using equation (\textcolor{black}{7}). For reference the seeing radius (or the full width half maximum radius) is shown with the black dashed line. Each panel also shows a ``background limit'' (grey dashed line), which is the radius at which the background begins to dominate the radial profile. The numbers correspond to the date of the composite image (column 1 of Table \ref{table:dust}).

We find no quantitative detection of dust through our radial profiles; that is, 2009 MS9 does not deviate significantly from a stellar radial profile. Even though activity was not detected, we can still use the error on the radial profile of 2009 MS9 to provide upper limit estimates of the dust production. Using the techniques described above, we measure the dust limits from the composites pre- and post-perihelion for 0.5, 2, and 10~$\micron$-sized grains. 

The maximum amount of dust we can detect is governed by where the background noise begins to dominate. We report the dust limit as the dust production rate where the surface-brightness of the object is equal to the sky noise (Table \ref{table:dust}). The uncertainty in this observation is defined as the change in dust production $\pm 1$ pixel from this value.

\subsection{Determining the Best-fit Albedo}
\label{sec:albedo}

Because of the uncertainty associated with the calculation of the H value from the MPC, we calculate our own estimate of the nucleus albedo using the radius obtained by WISE (11.5 km, section \ref{sec:WISE}) and photometric observations of 2009 MS9. 

We compute the nucleus contribution to the photometry assuming a \textcolor{black}{geometric albedo $p$} and radius, $R$, such that

\begin{equation}
\label{eq:nucleus}
pR^{2} = 2.235 \times 10^{22} r^{2}
\Delta^{2} 10^{[0.4(m_{\odot})-m]} 10^{0.4(\beta\alpha)}
\end{equation}

\noindent
where $r$ and $\Delta$ are the helio- and geocentric distances [au] and m($\odot$) and m are the apparent magnitudes of the sun and comet \citep{meech1986}. The term $\beta$ represents the simplified linear phase function, with phase angle, $\alpha$ [deg] and $\beta$ is a constant ranging from 0.02-0.04 mag / degree \citep{meech1986}. Thus the main variables that affect the overall brightness of the asteroid light curve are the \textcolor{black}{geometric} albedo and size, and the rest come directly from the object's orbital geometry.

For 2009 MS9, because its perihelion is at 10.99 au, the likely driver for activity is either CO or CO$_2$. \textcolor{black}{CO ice can lift observable dust grains as far out as 25 au, and CO2 ice sublimates inside of 12-13 au \citep{meech2004}, and this further depends on the evolution of the nucleus interior prior to the observed epoch.} \textcolor{black}{If we model the target as active throughout the entire duration of it's orbit with typical CO or CO2 sublimation (see section \ref{sec:sublimation}), then the WISE measurements on the nucleus size are just an upper-limit.} \textcolor{black}{A simpler, equally likely explanation could is to assume that 2009 MS9 is inactive pre-perihelion and active post-periheion, and this time delay is caused by CO or CO2 buried beneath the surface. For this scenario, fitting an asteroid curve to the observations pre-perihelion (e.g. for TA$<$0$^{\circ}$) will allow us to constrain the \textcolor{black}{geometric} albedo given the size of the target, which allows us to compare this to the calculated H-value from the MPC. We do so while keeping in mind the target's nucleus may indeed be an upper limit.}

We use Equation \ref{eq:nucleus} to calculate the nucleus light curve for varying albedos ranging from $0.15$ to $0.40$, in increments of 0.001. For each iteration, we calculate the sum of residuals of model from the observations, commonly known as the Chi-square statistic:

\begin{equation}
\chi^2 = \sum{\frac{(x_i-\mu)^2}{\sigma_i^2}}
\end{equation}

\noindent
where $x$ is the observed magnitude, $\mu$ is the expected magnitude from the asteroid model, and $\sigma$ is the error on each observed magnitude. Our desire is to minimize the Chi-Square statistic, or weighted residuals so that we obtain the best fit model to the observations. 

If the model is a good fit to the data, we expect a $\chi^2$ value approximately equal to our degrees of freedom, which is the difference between the number of observations and free parameters in the assumed model. Hence the Chi-square statistic is redefined as $\chi^{2} / \nu$, or the reduced Chi-square statistic, where good fits have a reduced Chi-square value of $\sim$1. 

For our model and data we have 1 free parameter and 29 observations before TA=0$^{\circ}$, which yields 28 degrees of freedom. To find the best fit model \textcolor{black}{we need to minimize the weighted difference between the observations and what the model predicts. To do so we need to also factor in the 0.4 mag rotation \citep{petit2017} into the interpretation of our analysis - this is governing what we consider the best fit model.}

This is much higher than any single observation's error, hence taking the individual photometric errors into account the reduced Chi-Square statistic will be abnormally large. \textcolor{black}{ If we add the error associated with rotation to the photometric error}, the $\chi^{2} / \nu$ will be lower than expected because each data point will not necessarily contribute 1-sigma to the statistic as expected. Hence the most valuable parameter for us is weighted residual itself, minimized at a specific albedo. 
With this radius the best fit light curve to the data has an albedo of 0.25 $\pm 0.04$, where the error is defined as the deviation from 0.25 at 2 times the weighted residual. Note that this is consistent (although slightly lower) with the albedo and error calculated from the MPC H value $0.4 \pm 0.14$.

\begin{figure}
\includegraphics[width=.47\textwidth]{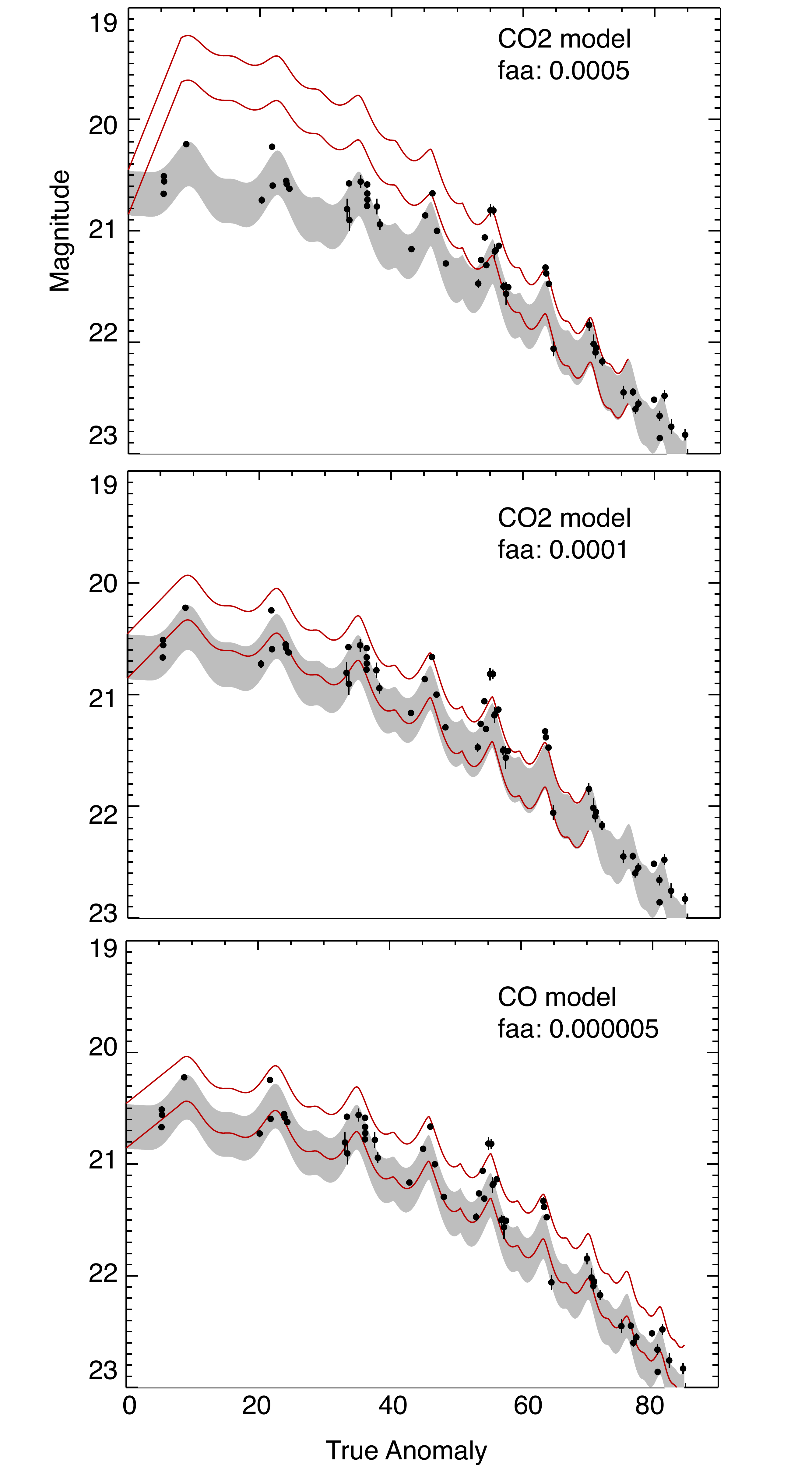}
\caption{Three sublimation models showing how the light curve changes between volatile species CO and CO$_{2}$ in comparison to photometry. {\bf Top}: CO$_{2}$ model matched to the brightnesses of the last data points of the target. Here the sublimation light curve does not match near-perihelion observations. {\bf Middle}: CO$_{2}$ model matched to the initial photometry of the target. Observations after TA$=50$ do not match sublimation model. {\bf Bottom}: CO sublimation matched to the photometry. This provides the best fit up to TA$=65$, however past this is too bright for observations.}
\label{fig:sublimation0}
\end{figure}

\subsection{Sublimation Modeling}
\label{sec:sublimation}

\begin{center}
\begin{table}
\centering
\caption{\label{tab:model}Sublimation Model Parameters}
\begin{tabular}{llll}
\hline
\hline
Parameter & M1 & M2 & M3 \\
\hline
{\bf Radius} [km] & 11.5 & -- & -- \\
Emissivity, $\epsilon$ & 0.9 & -- & --\\
Phase function [mag/deg], $\beta_{nuc}$ & 0.04 & -- & --\\
Phase function [mag/deg], $\beta_{coma}$ & 0.02 & -- & --\\
Grain Density [kg m$^{-3}$], $\rho_{grain}$ & 1000 & -- & --\\
Nucleus Density [kg m$^{-3}$], $\rho_{nuc}$ & 400 & -- & --\\
{\bf Albedo}, $p_{v}$ & 0.25 & -- & --\\
\hline
{\bf Grain size} [$\mu$m],$a_{grain}$ & 2 & 0.5 & 10 \\
{\bf Log Fract. Active Area} [\%],$f_{CO}$ & -5.9 & -5.3 & -4.5\\
\hline
\end{tabular}
\end{table}
\end{center}

\vspace{-.5cm}
We use an ice sublimation model that computes the amount of gas released due to solar heating and its overall effects on the object's heliocentric light curve \citep{meech1986,meech2004}. This ultimately results in an accurate prediction of gas loss and how the sublimation affects the object's brightness as it moves through its orbit (i.e. see \citet{snodgrass2013} for a comparison of this model to in-situ gas flux measurements).

The sublimation light curve is produced using a fixed aperture radius that combines the overall brightness from the nucleus with the coma, where the latter is produced by the gas sublimation and consequent drag of dust particles from the surface \citep{meech1986}. We then compare the sublimation light curve to the photometric data to model the suspected activity for our target. There are 10 model free parameters that are involved: ice type, nucleus radius, albedo, emissivity, grain density, nucleus density, dust size, phase function, thermal conductivity and fractional active area (FAA). Often preliminary observations can reveal constraints about many of these parameters, allowing for an accurate prediction of the sublimation associated with the observed data with adjustments to only a few of the variables.

\textcolor{black}{FAA is the fraction of the cometary surface that is sublimating. Besides modeling, this value can be obtained by measuring a comet’s total water production rate (from OH) and their radius (assuming that the nucleus is spherical) \citep{ahearn1995}. Typical FAAs for comets with water ice sublimation vary lie between 1-4\% \citep{ahearn1995}. Much higher FAAs have been observed for targets with large, sublimating icy-grains. An example of this is 103P/Hartley 2 which required over 100\% of the comet’s surface area to be actively sublimating \citep{lisse2009, groussin2004}. For hyper-volatile species like CO and CO2, FAAs are typically much lower than 1\% \citep{ahearn1995, meech2017}, although historically this has been difficult to accurately assess. This is because LPCs with hypervolatiles often have unknown nucleus sizes due to being active at large distances. Manx comets, or objects on LPC orbits with unexpectedly low to no activity, have even lower FAAs; for example 2013 LU28 is theorized to have a FAA of $10^{-4}$ and $10^{-6}$ for CO$_{2}$ and CO respectively \citep{slemp2022} and A/2018 V3 has a FAA of 10-5 for water sublimation \citep{piro2021}}

\textcolor{black}{We assume typical cometary values for emissivity \citep{meech2017}, nuclear and coma phase function \citep{krasnopolsky1987, meechjewitt1987}, nucleus density \citep{thomas2013a, thomas2013b, jorda2016}, and grain parameters such as size and density based on both ground-based comet data and the Rosetta mission \citep{yang2014, fulle2016}. While grain sizes follow a power-law size distributions from 0.1 $\mu$m to mm, the small particles are shown to dominate the emission \citep{fulle2016}. Thus for most models we adopt an average grain size of 2 microns, and vary this between submicron and 10 micron grains for selected CO models (see discussion below).} The parameters are summarized in the first seven rows of Table \ref{tab:model}. For this paper our derived nucleus size and albedo were used this to determine the fractional active area that best fits photometric observations. 

\begin{figure*}[hp!]
\includegraphics[width=1\textwidth]{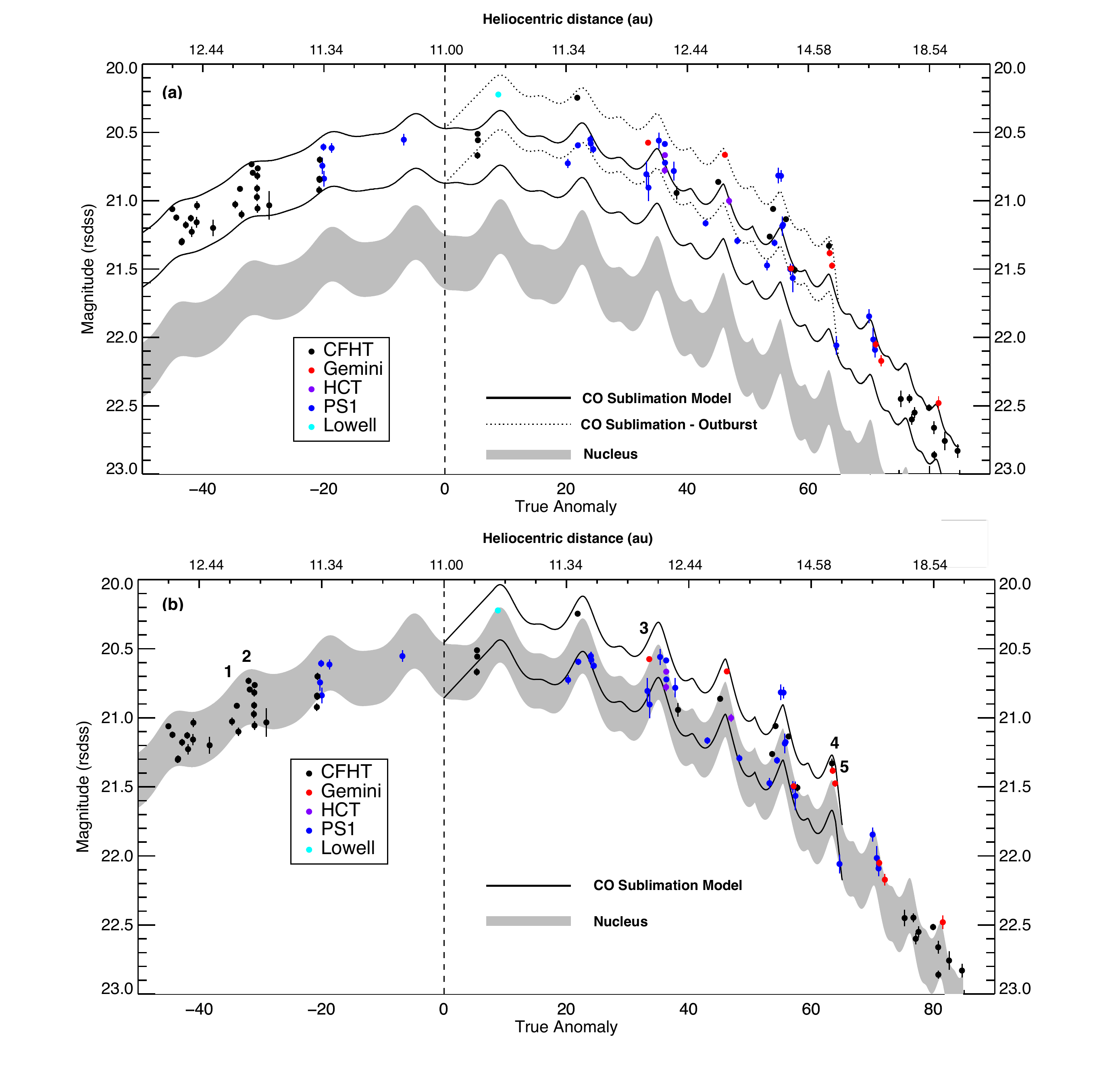}
\caption{\textcolor{black}{The photometric $r$-band light curve of 2009 MS9 as a function of its orbit assuming CO sublimation for two different activity scenarios. 2009 MS9's 0.4 magnitude variation encloses a range of magnitudes (grey-shaded region represents nucleus). (a) Modeling CO-sublimation pre- and post-perihelion. A baseline FAA of xxxx (black line) is needed plus a FAA of xxxx (dotted line) post-perihelion to match photometry. An arbitrary nucleus size of 8 km (grey) is shown. (b) CO sublimation (black line) post-perihelion with an FAA of $5 \times 10^{-6}$ assuming prior inactivity. Nucleus is 11.5 km (grey) from WISE.}} 
\label{fig:sublimation}
\end{figure*}

\textcolor{black}{First we explore the volatile species responsible for activity. Because of 2009 MS9's large perihelion, if it is active it has CO and/or CO$_{2}$ sublimation. We investigate the expected light curve based on both CO and CO$_{2}$ models. We construct the heliocentric light curve with observations from professional telescopes, taking into account that the 0.4 mag rotation creates an range/envelope of possible brightnesses for any given date (Figure \ref{fig:sublimation0}). Photometry is shown in rsdss magnitude and we take the average of the observations taken on the same observing night. We plot the expected contribution of the nucleus using equation 9 and the 0.4 mag rotation (grey-shaded region) knowing the size to be 11.5 km and with the estimated 0.25 albedo}

\textcolor{black}{We shows three different models in Figure \ref{fig:sublimation0}. CO$_{2}$ modeling (top and middle panels) requires a changing FAA to explain the behavior post perihelion. On the other hand, a CO model (bottom panel), while matching most data points eventually deviates from the data past TA$=65$. If we assume CO is the dominating volatile species, then the comet active stops beyond this. Activity dominated by CO or CO$_{2}$ is thus likely, however for this paper we adopt the simplest scenario which is that 2009 MS9 has CO sublimation that is quenched past TA$=65$. }

\textcolor{black}{Next, we investigate the duration of the suspected activity for 2009 MS9 with two CO models (Figure \ref{fig:sublimation}). In Section \ref{sec:albedo} we briefly mentioned the possibility that 2009 MS9 is active pre and post perihelion. If this is the case then the measured diameter (grey-shaded region) from WISE is an upper limit, and the target must have an outburst in activity post perihelion  (\ref{fig:sublimation}, panel a). For this, we arbitrarily pick a nucleus size of 8 km and show that with 2 micron-sized grains, a FAA of $2.5 \times 10^{-5}$ is needed pre-perihelion and $4 \times 10^{-5}$ post-perihelion.}

Figure \ref{fig:sublimation}, panel b shows the best fit model for only post-perihelion CO sublimation corresponding to an FAA of $5\times10^{-6}$ which best fits the data post perihelion up to TA=65$^{\circ}$, where the activity appears to cease. 

\textcolor{black}{From this point onward, we focus on the second scenario (Figure \ref{fig:sublimation}, panel b) that 2009 MS9 was inactive pre-perihelion and vary the grain sizes for our models. TNO 2009 MS9 is far enough from the Sun that any ejected grain will have the same residence timescale around the nucleus. Therefore it’s useful to model 2009 MS9’s activity assuming different dominating grain sizes. We thus report the CO model and resulting FAA with grain sizes 0.5,2, and 10 microns (Table \ref{tab:model}).}

\textcolor{black}{The sublimation model with input FAA outputs the gas production rate. Using a 1:1 ratio of gas and dust, we can turn this into a dust production rate. Next, we have independently derived the limit on the dust production rate using equation 8 and the surface brightness profiles of the composite images. This limit on dust production value must be higher than the dust production predicted by sublimation, otherwise we should be detecting the dust in the images. This is how the FAA can be constrained.}

If we assume a 1:1 ratio of gas and dust from the sublimation model, the dust mass loss rate (dm/dt) for this active fraction at the dates where we have dust limits from composites and where we assume the target to be active, we find that the dust mass loss rate is between \textcolor{black}{0.01-0.02 kg s$^{-1}$, 0.04-0.08 kg s$^{-1}$, and 0.2 - 0.5 kg s$^{-1}$ from 2013 to 2019 for 0.5,2, and 10 micron-sized grains} (Table \ref{table:dust}, ``Predicted Dust''). For all of the surface brightness profiles except for one, this is less than the upper limit on dust computed for each grain size. Thus 2009 MS9 may be active post-perihelion but the activity does not lift enough dust for a detectable coma in our composite images. The outlier surface brightness profile is from the composite image taken near perihelion on 2013 October 4, which has a more sensitive dust limit which suggests that for 2~$\micron$-sized grains we should have detected activity. \textcolor{black}{However, there are two factors that could have affected this result: 1. The dust-viewing geometry was poor (i.e. the dust was spread out over a larger area), and thus the dust may not have been concentrated enough for a high S/N for detection, and 2.} Since this data point is near perihelion (TA=8$^{\circ}$), the target may not have been active yet, as its photometry places this data just within the nucleus-only envelope. 

\textcolor{black}{In addition to typical surface ice sublimation presented above, another driver of activity is the H$_{2}$O amorphous ice to crystalline ice transition. Amorphous water ice forms in low temperatures and traps gas from more volatiles species in the process. When the ice anneals or undergoes a phase change to crystalline ice, this will release the trapped gas. \cite{meech2009} show that for a non-rotating body the amorphous-crystalline phase change occurs between 120-160 K starting as far out as 11.5 au. For rapid-rotating isothermal bodies, this begins at much closer distances, inward of 7 au \citep{meech2009}. Real comets are likely between these two cases. 2009 MS9's perihelion is 10.99 au, and is a rotating body, therfore it's likely too far out for amorphous-to-crystalline ice transition to be the cause of activity. Whether or not ice annealing (occurs as far out as 59 au for rapid-rotators) could be a driver of activity could be explored in a future paper \citep{meech2009}.}

\subsection{Thermal Modeling}
\begin{figure*}[t!]
\begin{center}
\includegraphics[width=1\textwidth]{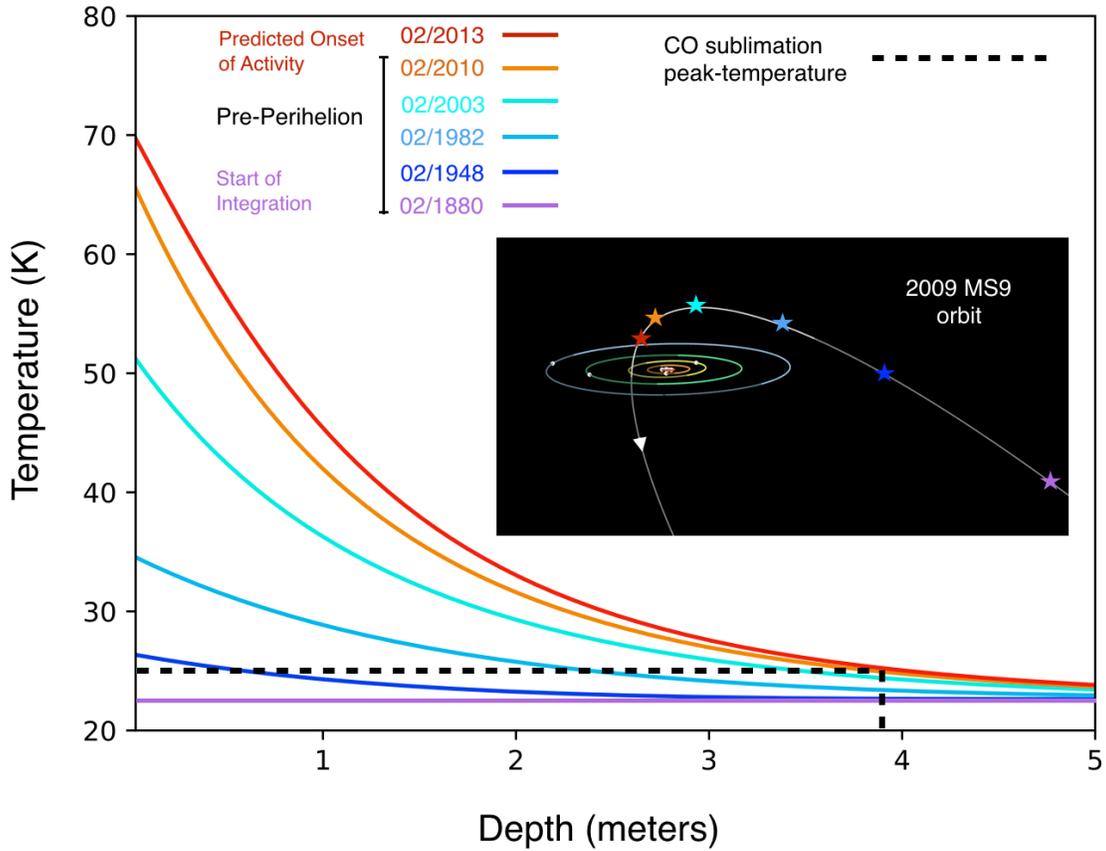}
\caption{The temperature profile as a function of depth for different positions (color coded in the top panel and matched to the stars of the same color-scheme in the second panel) the along the orbit of comet 2009 MS9. As 2009 MS9 gets closer to the sun, the temperature at its surface increases and this propagates to the interior, following 1D-heat conduction. We map the temperature profile of the near surface of the comet by integrating from 1880, where the object is too far ($~$130 au) to start CO sublimation, up until perihelion in 2013 with integration steps of one day.  At the object's perihelion is where we predict the onset of activity, and at this time the depth with the temperature corresponding to peak CO sublimation (25 K) is $\approx$ 3.9 m from the surface. }
\label{fig:therm}
\end{center}
\end{figure*}

The sublimation model described above accurately predicts the photometry produced by sublimation of volatiles at the surface, however, ice can be buried at a depth beneath the surface. This will result in a delay in the onset of activity depending on the thickness of the insulating layer and modeling this requires a different technique. 2009 MS9 is predicted to have become active near its perihelion, at 10.99 au, which is at a closer heliocentric distance than would be expected for CO ice sublimation on the surface, which occurs as far out as 130 AU, where the effective temperature at the surface would be 25 K at that distance \citep{meech2004}. 

It is also possible that CO$_2$ sublimation may lift optically detectable grains as far out as 20 au, however the light curve and active fraction suggest that the sublimation on 2009 MS9, if present, would be dominated by CO. Regardless of the volatile species, the simplest explanation for this delay in onset of activity is if the volatiles were buried at a depth beneath the surface. Using the delay of the onset of activity, we can solve for the depth using the 1D heat conduction equation subject to a time variable surface boundary temperature condition as the comet moves along its orbit. This method has been used for other comets to estimate the depth of volatiles and has been shown to agree with results obtained by more sophisticated thermal modelling \citep{meech2016}. 

The 1D heat equation is:

\begin{equation}
\frac{\partial T}{\partial t} = k \frac{\partial^{2} T}{\partial z^{2}}
\end{equation}

\noindent
where k W/cm $\times$ K is the thermal conductivity of the comet, and is represented by

\begin{equation}
    k = \frac{I^{2}}{\rho C_{p}}
\end{equation}

\noindent
where $\rho$ is nucleus density [kg/m$^{3}$], I [J K$^{-1}$ m$^{-2}$ s$^{-0.5}$] is the object's thermal inertia, and $c_{p}$ is the heat capacity [J kg$^{-1}$ K$^{-1}$]. The latter two have been determined for comet 67P/Churyumov-Gerasimenko by the $Rosetta$ mission where I has a range of values from $10 < I < 50$ J K$^{-1}$ m$^{-2}$ s$^{-0.5}$ and $c_{p} = 500$ J kg$^{-1}$ K$^{-1}$ \citep{ali-lagoa2015,gulkis2015} which are in agreement with measurements reported from the \textit{Deep Impact} and \textit{Stardust-NExT} missions \citep{groussin2013}. For simplicity we use a thermal inertia of $I = 10$ J K$^{-1}$ m$^{-2}$ s$^{-0.5}$ for our modelling.

To solve the 1D heat conduction equation, we use a Crank-Nicholson scheme, such that 
\begin{equation}
    \begin{split}
    -rT^{n+1}_{i+1} + (1+2r)T^{n+1}_{i} - rT^{n+1}_{i-1} = \\
    rT^{n}_{i+1} + (1-2r)T^{n}_{i} + rT^{n}_{i-1}
    \end{split}
\end{equation}

The value $r$ is a unitless numerical constant that is equal to $\frac{\Delta t}{1\rho c_{p} \Delta x^{2}}$, $i$ refers to the spatial step starting from the surface to the interior of the comet, and n refers to the time step of the equation. This is a second-order implicit numerical scheme that uses the central difference between spatial steps to numerically solve the equation for the thermal profile at the next time step. It is found to be unconditionally stable for values of $r$ less than 1. We use time steps of 0.25 days, or 6 hours, and spatial steps of 0.001 meters down to a depth of 100 meters for our calculation. 

The outer boundary condition of the temperature at the surface of the comet, T$_{eq}$, is a function of the comet's heliocentric distance. It is the effective temperature at the surface of the comet at its position along the orbit, and it is represented by:

\begin{equation}
    T_{eq} = T_{\odot}(1 - p_{v})^{\frac{1}{4}} \sqrt{\frac{R_{odot}}{2d}}
\end{equation}

Where $T_{\odot}$ is the temperature at the surface of the sun, $R_{\odot}$ is the radius of the sun, d is the heliocentric distance, and $p_{v}$ is the albedo of the comet. 

We start the numerical integration in 1880 (Fi. \ref{fig:therm}, purple), because at this time 2009 MS9 is at 130 au and this is colder than the temperature required for CO sublimation. We integrate until 2009 MS9 reaches its perihelion (Fi. \ref{fig:therm}, red), which is our predicted epoch for the onset of activity, to determine the depth that reaches the critical temperature for CO sublimation (Fi. \ref{fig:therm}, dashed line) (which is at 25 K.)

At 2009 MS9's perihelion on 2013 February 13 at 11 au, we find that the depth of the CO-ice layer that reaches the limit 25 K for CO sublimation is just $\sim$4 meters below the surface.

\subsection{Spectral Reflectivity}
\label{sec:refl}

The spectral reflectivity at each wavelength was calculated relative to the \textcolor{black}{gsdss} filter using the following equation:

\begin{equation}
R_{\lambda}=
\frac{10^{-0.4[m_{\lambda}-m_{\lambda}(\odot)]}}
     {10^{-0.4[m_g-m_g(\odot)]}}
\end{equation}

\begin{equation}
\sigma_{R\lambda}= 
0.92103\ R_{\lambda} \bigg[ \sigma_{\lambda}^{2}+\sigma_{g}^{2}+\sigma_{\lambda}(\odot)^{2}+\sigma_{g}(\odot)^{2}\bigg]^{0.5}
\end{equation}
In these equations $m_{\lambda}$ is the magnitude in the bandpass, $\sigma_{\lambda}$ is the error on $m_{\lambda}$, $\sigma_{r}$ is the error on $m_{r}$, and $m_{\lambda}(\odot)$ is the absolute magnitude of the Sun in that bandpass with error $\sigma_{\lambda}(\odot)$. We use $m_{g}(\odot)=5.12 \pm 0.02$, $m_{r}(\odot)=4.69 \pm 0.03$, $m_{i}(\odot)=4.57 \pm 0.03$, and $m_{z}(\odot)=4.60 \pm 0.03$.

The spectral slope, $S$, between two filters, normalized at filter $g$, is calculated with the following relation:

\begin{equation}
S_{g} = \frac{10^{5}}{\lambda - \lambda_{g}} (-1 + 10^{-0.4((m_{\lambda}-m_{g})-(m_{\lambda}(\odot)-m_{g}(\odot)))})
\end{equation}

\noindent
where $\lambda$ is in units of angstroms, and $S$ is in units of $\% / 10^{3}$\AA. Because there are more than two filters that make up 2009 MS9's spectral reflectivity, we measure the spectral slopes normalized to the g filter for all of the corresponding filters (ex: r,i,z,Y if applicable), and then calculate the average spectral slope for each observation date. We compared this method to the traditional method of calculating the spectral slope by fitting a least-squares line to the normalized reflectivity. The two techniques yield comparable values to within $1\% / 10^3$\AA. 

To compare our spectral gradients to TNOs and comets in the literature, we change our normalized filter to V using the following equation: 

\begin{equation}
S_{V} = \frac{S_{g}}{1+\frac{\lambda_{V}-\lambda_{g}}{10^{5}}S_{g}}
\end{equation}

We collected six \textcolor{black}{sets of }colors for 2009 MS9 using either Gemini or CFHT at different epochs of the target's orbit which are displayed in Table \ref{tab:colors}. Two are pre-perihelion and four are post perihelion. On nights where multiple color sets \textcolor{black}{(i.e. gri color sequences)} where taken (e.g. 7/15/2010,10/01-10/06/2010), we take the average color per night minus the magnitude variation \textcolor{black}{due to the 0.4 magnitude rotation \citep{petit2017}}. For post-perihelion observations \textcolor{black}{with Gemini} we correct the magnitudes for the 0.4 mag rotation by interweaving the g filter between other filters, fitting a linear model to the change in g magnitudes over time and subtracting this from the measured magnitudes. The change in g for each Gemini observation is consistent with the 0.4 magnitude difference observed by \citep{petit2017}. 

\begin{figure*}[ht!]
\includegraphics[width=1\textwidth]{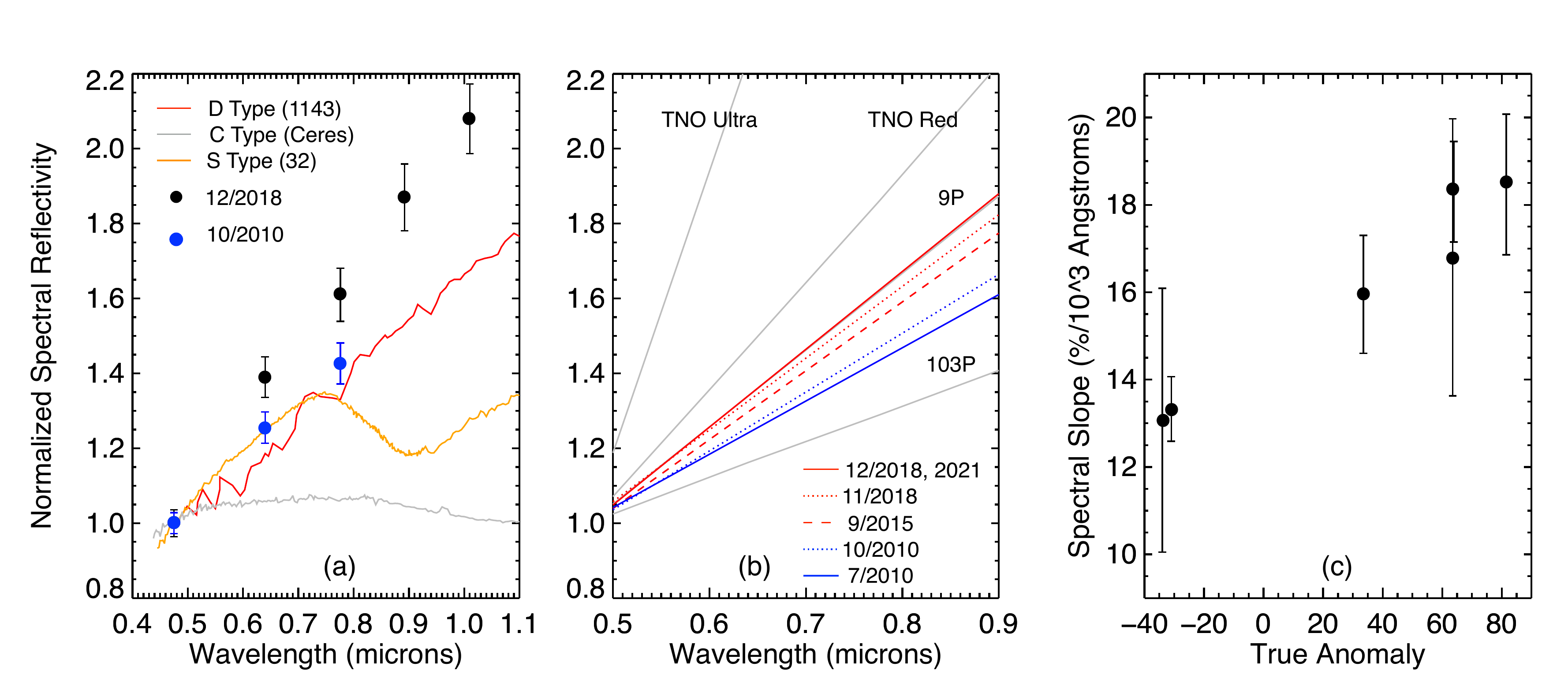}
\caption{
(a) 2009 MS9 colors pre- and post-perihelion in comparison to various SMASS-2 spectra types of other primitive bodies. The spectral reflectivity indicates a surface that is analogous to D-type asteroids. Pre-perihelion 2009 MS9's colors (blue) are bluer in comparison to its colors post perihelion (black). \textcolor{black}{(Note: The 12/2018 observation refers to observation taken on 11/30/2018. It is abbreviated to distinguish it from the observation earlier in the month taken 11/10/2018.)} (b) 2009 MS9's spectral slope has subtle changes over time that indicate redder colors post-perihelion, when we believe the target to be active. (c) comparing 2009 MS9's spectral slopes and their error to one another as a function of time suggest that the target is redder when active, however this results is not statistically significant. The numbers refer to the epoch of 2009 MS9's orbit the colors were measured which is shown on Fig. \ref{fig:sublimation}. 
}
\label{fig:reflectivity}
\end{figure*}

We compare 2009 MS9's spectral reflectivity to a few \textcolor{black}{SMASS-2 asteroid surfaces \citep{bus2002}} and find that like comets and other outer solar system bodies, its surface is similar to D-type asteroids in that \textcolor{black}{the slope of the data points  
is red and lacking a 1 micron absorption dip.} However, it appears to have a spectral slope that changes over time. Fig. \ref{fig:reflectivity}a shows the maximum \textcolor{black}{difference in spectral slope} between 2009 MS9 pre and post perihelion in comparison to asteroid types.

The orbital properties of 2009 MS9 put it in the unique position to straddle the boundary between different dynamical classes of minor bodies; specifically long period comets and trans Neptunian objects (Figure \ref{fig:reflectivity}b. Because of this, we compare 2009 MS9's changing colors to both types of objects. Pre-perihelion 2009 MS9 is closer to the colors of short period comet (SPC) 103P-Hartley \textcolor{black}{\citep{li2013}}. Post perihelion the colors appear redder, as the spectral slope increases to resemble that of short period comet 9P \textcolor{black}{\citep{li2007}}. The complete evolution of the 6 reflectivities is shown in Fig. \ref{fig:reflectivity}b in comparison to TNO and comet colors. In comparison to red and ultra-red TNOs, 2009 MS9 is considered ``grey'' (\textcolor{black}{\citealt{delsanti2004}}; see section \ref{sec:tnocolors} for further discussion). The corresponding spectral slope as a function of orbit in Fig. \ref{fig:reflectivity}c. There appears to be a linear increase in spectral slope as the 2009 MS9 moved throughout it's orbit.

\begin{table*}
\caption{\label{tab:colors} 2009 MS9 Colors and Spectral Slopes}
\begin{center}
\begin{tabular}{|cccccccc|}

\hline
Date & Telescope & g-r & r-i & i-Z & Z-Y & S$_{g}$ ($\%/10^{3}$ \AA) & S$_{V}$ ($\%/10^{3}$ \AA)\\
\hline
7/2010& CFHT & $0.64\pm0.02$ & $0.37\pm0.02$& -- & -- & $14.53 \pm 3.71$ & $13.07\pm 3.02$\\
10/2010 & CFHT &$0.66\pm0.01$  &$0.33\pm0.01$ & -- & --  & $14.82 \pm 0.92$ & $13.33\pm 0.74$ \\
\hline
\hline
9/2015 & Gemini&$0.74\pm0.02$  &$0.24\pm0.02$ & $0.27 \pm 0.02$ & -- &  $18.13 \pm 1.73$ & $15.95\pm1.35$\\
11/2018& Gemini&$0.77\pm0.03$  &$0.22\pm0.03$  & $0.34\pm0.03$ & $-0.33\pm0.05$ & $19.81 \pm 3.59$ & $16.80\pm3.17$\\
12/2018 &Gemini&$0.80\pm0.02$  &$0.27\pm0.02$  & $0.19\pm0.02$ & $-0.16\pm0.04$ & $21.22 \pm 1.56$ & $18.30\pm1.15$\\
12/2021 &Gemini&$0.798\pm0.03$  &$0.291\pm0.04$  & $0.13\pm0.04$ & -- & $21.46 \pm 2.17$ & $18.47\pm1.61$\\
\hline
\end{tabular}

\vspace{0.2cm}

\end{center}
\textcolor{black}{{\bf Note:} The table is split into observations pre- and post-perihelion indicated by double horizontal lines. The 12/2018 observation refers to the 11/30/2018 observation. It is abbreviated to distinguish it from the observation earlier in the month taken 11/10/2018.}
\end{table*}

%\clearpage

\vspace{0.5cm}
\section{Discussion}
\subsection{Colors In Comparison to TNOs}
\label{sec:tnocolors}

Trans-Neptunian Objects include minor solar system bodies orbiting the sun with a semi-major axis greater than that of Neptune, most of which belong to the Kuiper Belt. The Kuiper Belt has been found to have two major dynamical classes – the classical “cold” objects distributed near inclinations of 3 degrees and the dynamically excited, “hot” objects or scattered objects with a broad range of inclinations \citep{Brown2001}. These two populations are also distinguished by their colors. Studies have consistently shown that dynamically excited TNOs are less red then the cold, classical population, and that color is specifically correlated with inclination \citep{Tegler1998, Tegler2000, Tegler2005, Hainaut2012, Marsset2019}

The dichotomy of colors and orbital types for trans-Neptunian objects is seen as a precursor for the dichotomy of colors of centaurs, which are minor bodies that are thought to be the next evolutionary step after a TNO is perturbed out of the Kuiper belt and sent on shorter period, less stable planet-crossing orbits \citep{Tegler2016}. However one very important question remains: is this observed color and orbital dichotomy for TNOs and centaurs indicative of evolutionary change or does it represent a difference in origin and/or composition for these minor bodies?

Historically, \cite{Jewittluu2001} predicted that irradiation and cratering of the surface was responsible for the large distribution of TNO colors. However this was shown to be implausible by \cite{Tegler2003}, who was the first to statistically prove the TNO population exhibited a color bimodality and that their homogeneous surface colors pointed to grey impact craters on radiation reddened crusts to being an unlikely physical driver for color differences.

\cite{Wongbrown2017} theorized that the cause of the bimodal of color distribution observed in the Kuiper Belt stems from the retention of volatiles in some KBOs, whereas others do not because they migrated in the protoplanetary disk during the solar-system’s formation. They stipulated that KBOs that acquired volatiles at the surface became irradiated and thus redder, which could explain the split between grey and red surfaces. 

Our work shows that  2009 MS9, an extreme TNO, closest in orbital properties to the hot dynamically excited KBOs, has an albedo of 0.25 is comparable to the range of albedos observed for other TNOs \citep{jewitt2015}, and a ``grey'' surface that becomes red after sublimation. The change in spectral slope for 2009 MS9 (from 13 to 18 \% / $10^{3}$ \AA, Fig. \ref{fig:reflectivity}, Table \ref{tab:colors}) bring it closer to the observed split between ``grey'' dynamically excited KBOs and ``red'' cold classical KBOs  at $20\% / 10^3$ \AA \, observed by \cite{Marsset2019}.  

LPCs often change their orbital properties after sufficient close planetary passes. The same may be true for 2009 MS9, whose perihelion falls close to the outer 3 gas giants. To investigate this further future work must be done to simulate 2009 MS9's orbit to see if such passages exist. If they do, this plus the evidence for sublimation may eventually place it in the regime of ``red'' cold classical KBOs. Like \cite{Wongbrown2017}, we theorize that volatiles play a role in creating the two color populations observed in the Kuiper belt. However, we suggest that sublimation and orbital evolution are the driving factors for the two color classes of KBOs.

{\subsection{Evidence for Activity-driven Surface Evolution}}

\textcolor{black}{While it's been predicted that the color differences in the TNO population arise from the retention of volatiles from different formation environments \cite{Wongbrown2017}, we suspect color changes (reddening) of TNOs  may be sublimation driven based on the activity of our target. Specifically, the reddening of 2009 MS9 arises from the build-up of a dust mantle due at the surface of the comet.}

2009 MS9's colors are very similar to the spectral reflectivity of known Jupiter-family comets \citep{li2013,kelley2017}. However, 2009 Ms9's reddening post sublimation onset is contrary to what is seen for other comets when they become active. For example, the surface colors of 67P and C/2017 K2 became bluer as the comets approach perihelion and activity deposits fresh ice on the surface, which was observed in-situ for comet 67P \citep{meech2017, Fornasier2016}.  One explanation for 2009 MS9's reddening when it appears to be active is through the buildup of large grains brought to the surface from interior sublimation. When subsurface sublimation occurs, it has the potential to bring a variety of grains, including large (mm+) dust grains through the porous interior to the surface, where they may not be ejected into a dust tail because there is not enough force to lift them from the surface \textcolor{black}{or because the gas drag cannot accelerate them to a velocity higher than the escape velocity (and they fall back on the surface)} \citep{prialnik2004}. This process can create a dust mantle or crust that is redder than the prior surface due to the deposition of organic rich large grains that cover (or deplete) ice at the surface \citep{prialnik2004}. Additionally, large grains also have the potential to clog the comet porous medium, which may quench activity \citep{prialnik2004}. This may explain why 2009 MS9 had a delay in onset of activity, and why 2009 MS9 is no longer active now. 

\textcolor{black}{To investigate the concept of a dust mantle further - we have to consider the behavior of all grain sizes for 2009 MS9. Our sublimation model and dust model assumes one grain size for activity, however, previous studies of active bodies, including space-craft missions, suggest that there is a range of sub-micron to cm sized grains make up the nuclei (surface + subsurface) of small bodies that contribute to the dust tail during periods of activity \citep{meech2004, prialnik2004}.} \textcolor{black}{Due to the effects of scattering, grain sizes on the order of an observation's wavelength are most likely to be detected. Thus, in the optical, micron-sized grains should dominate in the observed dust tails of active bodies. The grain size-distribution in the dust tail follows a power law such that $f(r) \propto r^{n}$, where r is the radius of the grain.} \textcolor{black}{The power law index observed for previous short period comets is on average n$=-3.5$ \citep{fulle2004}. If n$>-3.5$ then this implies that the the brightness of the target from the dust tail is dominated by the scattering of the largest grains, even if observed in the optical \citep{fulle2004}.}

\textcolor{black}{The power index (whether measured as an average or time-dependent) is difficult to measure robustly for LPCs and other active bodies in the absence of in-situ space-craft missions. We might expect that the difference in formation environment and dynamical evolution between LPCs and SPCs may in turn cause different average indexes and size limits for each of these populations. For example, in Tony Farnham's thesis (1998), using a dust model that is described in \cite{Chu2020} he shows that dynamically new LPC comet Torres has a dust distribution that follows a power law with n=1.1 and further is unexpectedly devoid of micron-sized particles.}

\textcolor{black}{During the formation of a dust mantle, small grains (i.e. micron-sized) will move easier through the porous medium of the interior and are more likely to be ejected from the surface. This is because their velocity will be closer to the gas drag due to their higher acceleration in comparison to large grains.}

\textcolor{black}{So the question is: where are these micron-sized grains for 2009 MS9? There are two hypotheses for this.}
\begin{enumerate}
    \item \textcolor{black}{There aren't enough small grains to be detected in deep images}
    \item \textcolor{black}{The small grains simply aren't there.}
\end{enumerate}

\textcolor{black}{Both theories can be investigated with an order of magnitude approximation of what average dust production rate we would expect based on the window of observation for activity and the power law for the distribution of grain sizes. To a first approximation, to form a dust mantle the entire surface of the comet is covered in large grains that are small enough to travel through the porous interior - for example millimeter-sized. If we assume that the shape of 2009 MS9 and dust grains are spherical, then we would need $10^{14}$ mm-sized grains to cover the surface.} 

\textcolor{black}{Assuming a power law of n=$-3.5$ for typical SPCs, this means that $10^{23}$ micron-sized grains would be present and ejected from the surface. Next assume that each grain has a density of 2.2 g/m$^3$ - based on the average density of common materials on small bodies \citep{mukai1989, grun1990} and multiply this by the volume of one micron-sized spherical grain and the number of micron-sized grains to get the total mass of small-grains. If we assume that grains are emitted from the nucleus as early as 2009, our observation window to detect activity is 13 years. Dividing our total mass by this gives us an average dust production rate of $\approx 10$ kg/sec. This is higher than the limit we place on dust production using our surface brightness profiles and what our sublimation model predicts. If a power-law index of n$=-3.5$ is true, this suggests we should be detecting micron-sized dust. Since we are not, the micron-sized dust grains must not be there; perhaps they were ejected and dispersed before the first observation 13 years ago. }

\textcolor{black}{If we consider a higher index value for our power law, namely n=$-2.5$, this yields an average dust production of $0.06$ kg/sec for micron-sized grains. Note that this is consistent with what the sublimation model predicts, and is less than the limits that we have placed on the dust production rate for micron-sized grains using our surface-brightness profiles. This would suggest that the small dust is present, but does not have a high-enough dust-production rate to produce an observable dust tail in the images.}

\textcolor{black}{Adopting a power-law index of (-2.5) is compatible with theory 1, while a traditional index of -3.5 suggests theory 2 is more likely. Unfortunately, in order to ascertain a precise and accurate value of the power-law index we need to detect and model the dust, which is something we are not able to do at this time (but may be able to do with future JWST observations.) Nonetheless, the observation of our target's color changes (redenning), the outburst of sublimation post-perihelion suggesting volatiles at a depth, combined with either the absence of small grains or a power law $>-3.5$ suggests that the build up of a dust mantle as a result of cometary surface/subsurface evolution is highly likely. We would expect the volatiles to retreat further into the nucleus for future orbital iterations as the dust mantle builds up and the layer of ice depletes \citep{prialnik2004}.}

%\vspace{0.5cm}
\section{Summary}
\label{sec:discuss}
Recently astronomers have tried to directly detect activity in the Kuiper Belt as it is another abundant reservoir for icy bodies, however most KBOs are too faint for a coma to be observed. With newer, deeper surveys like Pan-STARRS and ATLAS we are able to study and discover these faint objects more efficiently, allowing us to examine volatiles over a larger range of distances. 2009 MS9 is one of these objects of interest, as it is classified as a TNO with a comet-like orbit that places its perihelion at 11 au where comets have been observed to become active. At these distances we can observe CO and CO$_2$ ice sublimation without the contamination of water ice, which activates at closer distances to the sun. This allows us to characterize activity of 2009 MS9 and other TNOs that have sublimation associated with more volatile species, which is important in understanding the distribution and roles of volatile transport during solar system formation and evolution. 

Our light curve constructed from a suite of professional telescopes is not compatible with an inert nucleus. We explore this further with composite images + radial flux profiles, and sublimation/thermal modeling. With this we set strong limits on the nucleus radius ($>11.5$ km) and geometric albedo (0.25) as well as cometary activity (dust + outgassing). The non-detection of dust in composite images suggests that large grains dominate activity. We confirm 2009 MS9 must be active with a hypervolatile species (CO or CO2) and this ice exists at a depth. The behavior of the light curve suggests CO as the dominant volatiles species. If active with CO, this activity is quenched near 15 au post-perihelion. 

Like other comets, 2009 MS9 exhibits colors analogous to D-type asteroids. When compared to other KBO colors, 2009 MS9 is considered ``grey''. However, the surface colors of our target changed throughout it's orbit. Pre- vs post-perihelion color evolution brings it closer to the observed boundary between ``grey'' and ``red'' KBOs. This suggests that sublimation may play an important role in creating the dichotomy of colors observed in the Kuiper Belt. Lastly, the timing of the color change combined with the proposed activity and presence of large grains suggests the accumulation of a dust mantle on the surface of 2009 MS9.

With JWST we will have the opportunity to constrain the colors of the target through spectra. 2009 MS9 will remain observable by large telescopes such as Gemini, CFHT, until it reaches 24th magnitude in 2027, however may be observable by LSST and JWST until the object reaches 27th magnitude in 2040, which is the limiting magnitude of the stacked images of LSST.

\section{Acknowledgements}
KJM and JTK acknowledge support through awards from the National Science Foundation AST1413736 and AST1617015. 

Data were acquired using the PS1 System operated by the PS1 Science Consortium (PS1SC) and its member institutions. The PS1 Surveys have been made possible by contributions from PS1SC member Institutions and NASA through Grant NNX08AR22G, the NSF under Grant No. AST-123886, the Univ. of MD, and Eotvos Lorand Univ.

Based also in part on observations obtained with MegaPrime/MegaCam (a joint project of CFHT and CEA/DAPNIA, at the Canada-France-Hawaii Telescope which is operated by the National Research Council of Canada, the Institute National des Science de l'Univers of the Centre National de la Recherche Scientifique of France, and the University of Hawai`i).

Additionally, observations were obtained at the international Gemini Observatory, a program of NSF’s NOIRLab, which is managed by the Association of Universities for Research in Astronomy (AURA) under a cooperative agreement with the National Science Foundation. on behalf of the Gemini Observatory partnership: the National Science Foundation (United States), National Research Council (Canada), Agencia Nacional de Investigaci\'{o}n y Desarrollo (Chile), Ministerio de Ciencia, Tecnolog\'{i}a e Innovaci\'{o}n (Argentina), Minist\'{e}rio da Ci\^{e}ncia, Tecnologia, Inova\c{c}\~{o}es e Comunica\c{c}\~{o}es (Brazil), and Korea Astronomy and Space Science Institute (Republic of Korea). This work was enabled by observations made from the Gemini North telescope, located within the Maunakea Science Reserve and adjacent to the summit of Maunakea. We are grateful for the privilege of observing the Universe from a place that is unique in both its astronomical quality and its cultural significance.

This publication also makes use of data products from the Near-Earth Object Wide-field Infrared Survey Explorer (NEOWISE), which is a joint project of the Jet Propulsion Laboratory/California Institute of Technology and the University of Arizona. NEOWISE is funded by the National Aeronautics and Space Administration.

We thank the staff of IAO, Hanle and CREST, Hosakote, that made these observations possible.  IAO and CREST are operated by the Indian Institute for Astrophysics, Bangalore.

This research has made use of the scientific software at www.comet-toolbox.com
(Vincent, J.-B., Comet-toolbox: numerical simulations of cometary dust tails in your browser, Asteroids Comets Meteors conference, 2014, Helsinki).

\bibliographystyle{aasjournal}
\bibliography{references}
\clearpage

\end{document}